\documentclass[twocolumn,a4paper,superscriptaddress,floatfix,longbibliography]{revtex4-2}

\usepackage[normalem]{ulem}
\usepackage{amsmath,amssymb,epsfig,color,bm,graphicx,alltt}
\usepackage{amsfonts}
\usepackage{latexsym}
\usepackage{wrapfig}
\usepackage{hyperref}

\newcommand{\I}{{\rm i}}

\begin{document}

\title{Itinerant electron metamagnetism for lattices with van Hove density-of-states singularities near the Fermi level}

\author{F.~A.~Vasilevskiy}
\affiliation{M. N. Mikheev Institute of Metal Physics, Yekaterinburg, Kovalevskaya street, 18, 620108, Russia}

\author{P.~A.~Igoshev}
\affiliation{M. N. Mikheev Institute of Metal Physics, Yekaterinburg, Kovalevskaya street, 18, 620108, Russia}
\affiliation{Ural Federal University, Russia, Yekaterinburg, Mira street, 19, 620002, Russia}

\author{V.~Yu.~Irkhin}
\affiliation{M. N. Mikheev Institute of Metal Physics, Yekaterinburg, Kovalevskaya street, 18, 620108, Russia}
\affiliation{Ural Federal University, Russia, Yekaterinburg, Mira street, 19, 620002, Russia}

\begin{abstract}   
Itinerant-electron metamagnetism 
is investigated within the Hubbard model for various lattices having van Hove singularities (vHS) in the electronic spectrum: face-centered cubic and orthorhombic lattices. The remarkable itinerant-electron metamagnetic transition occurs provided that the Fermi level is in the region with a strong positive curvature of the density of electron states typically positioned  between two close van Hove singularities. 
Orthorhombic distortion of a~tetragonal lattice is a promising mechanism for generating two closely split vHS with strong density-of-states curvature between them. 
A phase diagram in terms of electron filling and Hubbard interaction parameter is presented, which shows the paramagnetic-metamagnetic-ferromagnetic phase transition and regions of saturated and non-saturated magnetism. The standard Landau theory expansion based on~the~electron density of states in the vicinity of the Fermi level is demonstrated to be insufficient to describe the whole magnetic phase diagram including the itinerant-electron metamagnetic transition. 
\end{abstract}

\maketitle

\section{Introduction}
The study of magnetic phase transitions is important for a fundamental understanding of the nature of magnetism. One of the phase transition kinds is the phenomenon of itinerant-electron metamagnetism~(IEMM), which is a transition from a low-magnetization state to high-magnetization state under  application of a~magnetic field. Although itinerant metamagnetism has been discovered for a long time, \cite{wohlfarth1962collective}, a complete comprehension of this phenomenon has not yet been achieved.

Experimentally, the IEMM transition was for the first time detected in a pyrite-structure compound CoS$_2$~\cite{adachi1979further, mushnikov2002itinerant,goto1997magnetic}. CoS$_2$ is an itinerant ferromagnet with the Curie temperature $T_{\rm c}=122$ K. There are two possible ways towards appearance of IEMM transition in this compound. The first one is the substitution of S by~Se: due to broadening of the $3d$-band with increasing Se concentration, the density of states~(DOS) at the Fermi level decreases and the magnetic ordering changes. At doping values of $0.12 \le x \le 0.4$, Co(S$_{1-x}$Se$_x$)$_2$ becomes an exchange-enhanced Pauli paramagnet where IEMM transition in high magnetic fields $B_{\rm c}<20$~T is possible, with the critical magnetic field being smaller for smaller $x$ \cite{adachi1979further,goto1997magnetic} and  $B_{\rm c}$ being proportional to $T^2$. The temperature dependence of magnetic susceptibility typical for metamagnetic materials is observed --- a very high value of susceptibility at low temperatures as compared to Pauli paramagnets and a broad maximum at a certain temperature $T_{\rm max}$ (for Co(S$_{1-x}$Se$_x$)$_2$, this is $80$~K). Using the~spin fluctuation theory, it has been shown that at $T_{\rm max}$ there is a sign change of the fourth-order coefficient in the Landau expansion~\cite{Yamada1993}; thereby~$T_{\rm max}$ corresponds to the temperature at which metamagnetism vanishes. The second way is also related to the broadening of the $3d$-band due to the application of external pressure \cite{ mushnikov2002itinerant,goto1997magnetic}. With an~increasing pressure, the Curie temperature for CoS$_2$ decreases and the ferromagnetic~(FM)---paramagnetic phase transition changes its order from the second to the first one at $p\approx 0.4$~GPa. At pressure $p>0.4$~GPa, a metamagnetic transition above $T_{\rm c}$ is observed. 
The replacement S by Se results in decreasing  critical pressure of this phase transition.

The next class of relevant compounds is presented by~intermetallic compounds   $R$Co$_2$ with rare earth $R$=Lu, Y. YCo$_2$ and LuCo$_2$ are typical exchange-enhanced Pauli paramagnets. The IEMM transition in them was first predicted theoretically on the basis of density functional calculations for the YCo$_2$ compound, the critical magnetic field value $H_{\rm c}=900$~kOe~\cite{Yamada1985MetamagneticTO} being obtained, i.e., the transition occurs in strong magnetic fields. This was confirmed experimentally: the itinerant metamagnetism of compounds YCo$_2$ and LuCo$_2$ was found in strong magnetic fields $H_{\rm c}=690$ and $740$~kOe, respectively~\cite{goto1989itinerant,goto1990itinerant,mushnikov1999itinerant}. In these compounds, an appearance of the IEMM transition is due to the hybridization of the low-energy $3d$-band of cobalt with a high DOS and  high-energy $4d(5d)$-band of $R$ with a relatively low DOS, resulting in a region with positive DOS curvature at the Fermi level for resulting spectrum. 
%{[\blue??? Test this:]}The inclusion of a magnetic field increases the density of states at the Fermi level and a field-induced metamagnetic transition to the ferromagnetic state occurs when the Stoner criterion is fulfilled. 

The replacement of cobalt by aluminum in these compounds results in decreasing  the critical magnetic field value~\cite{sakakibara1986itinerant,aleksandryan1985metamagnetism, mushnikov1999itinerant,gabelko1987onset,fukamichi2001magnetic}. 
%\sout{This is due to the fact that at such a substitution the concentration of $3d$ electrons becomes smaller, due to which the Fermi energy decreases and the density of states increases.}
This is due to the fact that such a substitution results in decreasing  $3d$-electron filling, which in turn shifts the Fermi level with the corresponding increase of DOS. Also, such a replacement leads to an increase in the volume of  unit cell due to that the bandwidth narrows and DOS at the Fermi level increases. In the compound Y(Co$_{1-x}$Al$_x$)$_2$, IEMM transition is observed at $x<0.12$, whereas at $0.12<x<0.20$  weak ferromagnetism occurs \cite{yoshimura1985new}. 
The temperature dependence of magnetic susceptibility has a broad maximum at $T_{\rm max}$, typical for itinerant metamagnets.  Applying pressure results in an increase of~$H_{\rm c}$  due to~a~bandwidth increase~\cite{mushnikov1999itinerant}. 
%{\blue ???\cite{lemaire1966magnetic, Bloch1971TUDEDC}}.

A van Hove singularity in $3d$ electron band is a possible origin of the low-spin --- high spin ferromagnetic state transition in the compound LuCo$_3$~\cite{2020:LuCo3:Neznahin,2022:LuCo3:Radzivonchik}. 

A simple model of electron spectrum within nearest and next-nearest neighbor hopping approximation allows to describe the magnetic properties and itinerant metamagnetic transition for double-layer and triple-layer  ruthenates Sr$_3$Ru$_2$O$_7$~\cite{2003:Perry:Sr3Ru2O7} and Sr$_4$Ru$_3$O$_{10}$~\cite{2003:Cao:Sr4Ru3O10}.   
These results suggest that an~IEMM transition may have a single sufficient reason --- strong positive curvature of DOS in the paramagnetic phase produced by vHS in the vicinity of Fermi level. 

A  typical $5f$ itinerant metamagnet is UCoAl. 
%Paramagnetic in the ground state. 
In magnetic fields of about $0.6$~T directed along the $c$ axis, this compound undergoes a metamagnetic transition~\cite{mushnikov1999magnetic, mushnikov2002effect}. A growth of critical magnetic field with increasing pressure and temperature dependence of the susceptibility, typical for itinerant metamagnets,  are observed. For UCo$_{0.98}$Fe$_{0.02}$Al, ferromagnetism disappears at a pressure of $0.4$~GPa manifesting an IEMM transition: further pressure increasing results in 
linear increase of the critical magnetic field and gradual decrease of magnetization jump $\Delta m$. IEMM transition appears with increasing temperature, that is, there is a temperature-induced phase transition from the ferromagnetic state to the magnetic field-induced ferromagnetic state, predicted theoretically ~\cite{yamada2001itinerant} and observed for $3d$-band magnets~\cite{goto1997magnetic}.  The critical magnetic field grows almost linearly in~$T^2$ at any pressure.  

URhGe and UCoGe exhibit ferromagnetic and superconducting phase transitions at ambient pressure~\cite{aoki2019review, aoki2001coexistence, huy2007superconductivity}. Features of FM transitions in these compounds are strongly dependent on the magnetic field $H$ direction relative to the easy  magnetization axis. In URhGe, in the case $H \parallel b$ at $H=H_{\rm R}=120$~kOe (the spin reorientation field) the easy magnetization axis switches from the $c$ to $b$-axis, and a metamagnetic transition appears. Nuclear magnetic resonance spectra show the presence of phase separation in the magnetic field $H = H_{\rm R}$. The quantum critical end point of ferromagnetic (FM) wing structure appears due to the magnetic-field direction tilting from $b$ to $c$ axis around the $H_{\rm R}$ and is associated with field-reentrant superconductivity. 
Thermoelectric power measurements indicate that a Lifshitz transition occurs in a narrow window through $H_{\rm R}$. 
%Quantum oscillation and TEP experiments indicate that the Lifshitz transition is coupled to the variation of the FM fluctuation.
The compound UIr$_2$Si$_2$, being ordered antiferromagnetically at temperature below 5.5~K. exhibits metamagnetic transition occuring at~1.8~K in a magnetic field of 15.2~kOe. 
Despite the fact that this transition has a spin‑flop character, this compound  possesses some itinerant properties~\cite{2023:Szlawska:UIr2Si2}, which makes it close to itinerant metamagnets. 

Density functional theory (DFT) calculations suggests that there is a set of  rather weak vHS of density of states of UPt$_3$ in the vicinity of Fermi level. 
These features invoke both Lifshitz transition and non-linear dependence of magnetization on magnetic field, which is similar to an~itinerant metamagnetic transition at magnetic field 20~T~\cite{2021:McCollam:UPt3}.

In the ferromagnetic La(Fe$_x$Si$_{1-x}$)$_{13}$ at $x\geq 0.87$, a IEMM transition is observed at temperatures above the Curie temperature $T_{\rm c}$~\cite{fujita1999itinerant,fujita1999giant,fujita2003large}. At $x=0.88$, the site magnetic moment arising from the metamagnetic transition exceeds $1 \mu_{\rm B}$ and leads to a huge positive bulk magnetostriction of about $1.5 \%$ just above $T_{\rm c}$ \cite{fujita1999itinerant}. The huge magnetostriction in this system is due to a huge volume change caused by the IEMM transition. The critical magnetic field at $T_{\rm c}$ is zero and increases with increasing temperature, so for $x=0.88$ at $T_{\rm c} = 195$~K the transition magnetic field near room temperature is larger than a few tens of T. %{\blue [Povtor:] The compounds La(Fe$_{0.88}$Si$_{0.12}$)$_{13}$ exhibit  just above the Curie temperature is large enough to use them as a magnetostrictive material} \cite{fujita1999giant}.

In addition, the phenomenon of itinerant metamagnetism often accompanies the giant magnetocaloric effect, e.g., in MnAs \cite{gama2004pressure}, Gd$_5$(Si$_{1-x}$Ge$_x$)$_4$ \cite{tegus2002magnetic,pecharsky1997} and MnFeP$_{1-x}$As$_x$ \cite{tegus2002magnetic}.

UGe$_2$ \cite{taufour2010tricritical, aoki2019review,saxena2000superconductivity} and ZrZn$_2$ \cite{kimura2004haas, uhlarz2004quantum} are itinerant ferromagnets exhibiting a first-order transition at low temperatures under pressure. %The Curie temperature 
$T_{\rm c}$ decreasing under pressure $p$ and disappears at the critical pressure $p_{\rm c}$. 
From thermal expansion measurements, it follows that at low pressure the transition is of a second order, but near $p_{\rm c}$ it is of a first order. Magnetic measurements for different pressures indicate first-order magnetic phase transition corresponding to~a~magnetization jump at a critical pressure from a low-pressure FM$_2$ phase with  large magnetization to a high-pressure FM$_1$ phase with lower magnetization. 
%This FM$_1$---FM$_2$ transition is of a first order. 
Just above $p_{\rm c}$ a metamagnetic transition from the paramagnetic (PM) to the FM$_1$ state is observed at critical magnetic field $H_{\rm c}$. The metamagnetic transition inside the FM phase observed in UGe$_2$ %and ZrZn$_2$ 
was explained in Ref.~\cite{sandeman2003ferromagnetic} by a double-peak structure of the electronic DOS. This particular DOS structure can also provide an explanation for the FM$_1$---FM$_2$ transition.

Summarizing the experimental data, we can highlight the main features of itinerant metamagnets. 

1) The magnetic susceptibility has a broad maximum near the temperature $T_{\rm max}$. 

2) The value of the critical magnetic field $H_{\rm c}$ of~a~IEMM transition is proportional to $T^2$ at low temperatures. 

3) IEMM transition characteristics are very sensitive to external pressure and doping.

On one hand, itinerant metamagnetic transition %, which is a sharp change in magnetization when the magnetic field changes, 
can occur in an electronic system where magnetism has an itinerant nature. On the other hand, for the occurrence of a metamagnetic transition, it is necessary that the electronic structure of the system has significant features of the DOS: the condition for the occurrence of metamagnetism is a positive DOS curvature   (${d^{2}\rho}/{d\epsilon^2}>0$) at the Fermi level \cite{shimizu1982itinerant,levitin1988itinerant}. One of such mechanisms can be  presence of van Hove singularities in the electronic spectrum \cite{2022:Igoshev:vHS_fcc}. Modern DFT calculation methods allow to calculate the density of states in the Kohn-Sham spectrum, which can be used as a starting point for further studies, including simple mean-field approaches or advanced approaches taking into account for magnetic fluctuations and correlation effects.

The papers \cite{wysokinski2015criticalities,sandeman2003ferromagnetic,berridge2011role, yamase2023ferromagnetic} discuss the microscopic approach to the study of IEMM. In Ref.~\onlinecite{igoshev2023ferromagnetic}  the onset of a~ferromagnetic ordering in the~nondegenerate Hubbard model on a face-centered cubic~(FCC) lattice within the framework of the functional renormalization group method is investigated; the hopping integral configuration was taken to provide a giant van Hove singularity in DOS. In particular, it was shown that correlation effects can be taken into account for three-dimensional lattices in the framework of the Stoner theory by renormalizing the electron-electron interaction parameter $U$, which takes into account its screening in the partial-particle channel. 
Magnetic phase separation for FM-PM phase transition was considered within the Landau theory generated by giant vHS in FCC lattice~\cite{2024:Igoshev_MCE}. 

The aim of the present work is to study the metamagnetic phase transition in metals having van Hove singularities in the electronic spectrum. 
We will demonstrate that a crucial feature favoring IEMM is a large value of DOS curvature, which is just provided by one or two vHS.
The following lattices are considered: 
(1)~a square lattice with an anisotropic spectrum (rectangular lattice) in the nearest-neighbor and next-nearest-neighbor approximation, where the transfer integrals between nearest neighbors depend on the bond direction, which provides a~pair of~DOS peaks and a region between them with large curvature; 
(2)~an orthorhombic lattice obtained from a rectangular lattice by adding a transfer along the $z$-axis, which leads to a change of the two-peak structure of DOS to two-plateau structure and preservation of the region with positive curvature $\rho(\varepsilon)$; 
(3) FCC lattice with small $\tau$ ($\tau=t/t'$, $t$~$(t')$ being nearest-neighbor (next-nearest-neighbor) hopping integral); 
(4) FCC lattice with $\tau$ in the vicinity of $-1/2$~($\tau = - 0.52$ and $ - 0.54$) has a  DOS singularity in the form of van Hove plateau, corresponding to giant vHS (at $\tau = -1/2$), and a considerable DOS curvature~\cite{2022:Igoshev:vHS_fcc,2024:Igoshev_MCE,1998:Ulmke}.

\section{Electron spectrum and Hartree-Fock approximation  equations}\label{sec:dos}

The non-degenerate Hubbard model Hamiltonian reads
\begin{equation}\label{eq:Hamilt}
    \mathcal{H} = \sum_{ij\sigma}t_{ij}a^\dag_{i\sigma}a^{}_{j\sigma} + U\sum_{i}n_{i\uparrow}n_{i\downarrow} - h\sum_{i\sigma}\sigma n_{i\sigma},
\end{equation}
where $t_{ij}$ is hopping integral between sites $i$ and $j$, $U$ is a Coulomb interaction parameter, $h = \mu_{\rm B} H$ is magnetic field in energy units with $\mu_{\rm B}$ being the Bohr magneton, $a^{\dag}_{i\sigma}, a^{}_{i\sigma}$ are the creation and annihilation electron operators, $n_{i\sigma}=a_{i\sigma}^\dag a_{i\sigma}$ is the electron number  operator at lattice site $i$ with $z$ axis spin projection $\sigma = \pm1$. 

Below, we study the thermodynamics of the IEMM transition in the Hartree-Fock approximation~(HFA). The free energy in this approximation has the form ~\cite{Moriya_book}
\begin{equation}\label{eq:F_HFA}
    F_{\rm HF}(T, n, h| m) = F_0(T, n | m) + (U/4)(n^2 - m^2) - hm,
\end{equation}
where the free energy of the electron gas 
\begin{equation}
F_0(T, n, m) = \Omega_0(T, E_{\rm F}, \Delta) + m\cdot\Delta + n\cdot E_{\rm F}     
\end{equation}
is defined through the Legendre transformation of the grand potential for free electron gas,  
\begin{equation}
\Omega_0(T, E_{\rm F}, \Delta) = -T\sum_{\mathbf{k}\sigma}\ln\left(
1 + \exp(-\beta(\epsilon_{\mathbf{k}} - \sigma\Delta - E_{\rm F}))
\right),
\end{equation}
where $\beta = 1/T$ (temperature $T$ is taken in energy units), the Fermi level $E_{\rm F} = E_{\rm F}(n,m)$  and spin subband splitting $\Delta = \Delta(n,m)$  are determined from the equations
\begin{eqnarray}
\frac{\partial\Omega_0}{\partial E_{\rm F}} + n &=& 0, \\
\frac{\partial\Omega_0}{\partial \Delta} + m &=& 0.    
\end{eqnarray}
We rewrite these equations as the equations
\begin{eqnarray}
\label{eq:main_eqiations:n}
n = \sum\limits_{\sigma} \int d\epsilon \rho(\epsilon)f_\mu\left(  \epsilon +\dfrac{Un}{2}-\sigma \Delta\right), \\   
\label{eq:main_eqiations:m}
m = \sum \limits_{\sigma} \sigma \int d\epsilon \rho(\epsilon) f_\mu\left(\epsilon +\dfrac{Un}{2}-\sigma \Delta\right)         
\end{eqnarray}
for the site magnetization $m$ in units of $\mu_{\rm B}$ and chemical potential $\mu$ for given values of $U$, filling $n$, and magnetic field $h$. 
Here $f_{\mu}(E)= (\exp((E-\mu)/T)+1)^{-1}$ is the Fermi function, $\mu = E_{\rm F} + Un/2$, $\Delta = Um/2+h$, $\rho(\epsilon)$ is DOS for an electron spectrum $\epsilon_{\mathbf{k}} = (1/N)\sum_{ij}t_{ij}\exp[\I\mathbf{k}(\mathbf{R}_i - \mathbf{R}_j)]$
\begin{equation}\label{eq:DOS_def}
    \rho(\epsilon) = \frac1{N}\sum_{\mathbf{k}}\delta(\epsilon - \epsilon_{\mathbf{k}}),
\end{equation}
where $w_1 < \epsilon < w_2$, $w_{1,2}$ being band boundaries. 
Main equations \eqref{eq:main_eqiations:n} and \eqref{eq:main_eqiations:m} have, generally speaking,  several solutions
\begin{equation}\label{eq:state_solution}
    m = m_i(T,n,h),
\end{equation}
indexed by $i$. If more then one solution is obtained, the solution with the lowest free energy (\ref{eq:F_HFA}) is chosen.
The IEMM transition point corresponds to the fact that  two different solutions Eq.~(\ref{eq:state_solution}) have equal free energy at some magnetic field $h > 0$:
\begin{equation}\label{eq:metamagnet}
    F_{\rm HF}(T,n,h|m_1) = F_{\rm HF}(T,n,h|m_2).
\end{equation}
Equation (\ref{eq:metamagnet}) can be viewed as an equation on the critical magnetic field $h = h_{\rm c}(T,n)$. 
A saturated FM state is realized when effective Fermi level of lower spin subband $E_{\rm F} - (Um/2 + h)$ is positioned below $w_1$. 
%From Eqs.~(\ref{eq:main_eqiations:n}-\ref{eq:main_eqiations:m}) one gets the condition of saturated FM state %($n_\uparrow = n$, $n_\downarrow = 0$) 
%stability at $T = 0$  for given $n$ 
%\begin{equation}\label{eq:saturated_state}
%    n < \int\limits_{w_1}^{w_1 + Un + h}\rho(\epsilon)d\epsilon.
%\end{equation}

Note that the well-known Stoner criterion
\begin{equation}\label{eq:Stoner}
    U\rho(E_{\rm F}) > 1
\end{equation}
corresponds to the condition of existence of the local maximum of the  ground-state free energy $F_{\rm HF}(T,n,h = 0|m)$ at $m = 0$ in zero magnetic field, see Eq.~(\ref{eq:F_HFA}). 
Thus, the fullfillment of the Stoner criterion has an exact meaning of the impossibility of a paramagnetic (PM) state even as a metastable one. 

The spectrum of \textit{rectangular} (anisotropic square) lattice in the nearest- and next-nearest-neighbor approximation has the form 
\begin{eqnarray}\label{eq:e_sq_a}
    \epsilon^{\rm R}_\mathbf{k}(\tau_{\rm a}, \tau)=-2t[\cos k_x + \tau_{\rm a} \cos k_y + 2 \tau \cos k_x \cos k_y],
\end{eqnarray}
where $\tau_{\rm a} = t_{\perp}/t$, $\tau=t'/ t$, $t$ is the hopping integral along $x$, $t_{\perp}$ along $y$, $t'$  along the diagonal. 
Without loss of generality, we assume $\tau_{\rm a} < 1$. 
Here and below lattice parameter is taken as unity. 
%At $\tau_{\rm a} = 1$, the spectrum \eqref{eq:e_sq_a} reduces to the spectrum of the square lattice~\cite{2017:Igoshev_MCE}.  

%By analyzing the spectrum of the rectangular lattice, see~Eq.~(\ref{eq:e_sq_a}), the integral defining the DOS, see Eq.~(\ref{eq:DOS_def}),
%reduces to full elliptic integral of the first kind. 
The spectrum \eqref{eq:e_sq_a} produces DOS of rectangular lattice with bandwidth  $W_{\rm R} = ( {\rm max}[w_1, E_2] - E_{\rm min})$ by a direct calculation:
\begin{itemize}
\item $\tau < \tau_{\rm a}/2$: 
\begin{multline} \label{eq:rho_tau1}
\rho_{\rm R}(\varepsilon, \tau_{\rm a}, \tau) = \frac2{\pi^2t}\frac1{\sqrt{|(1 - \tau_{\rm a})^2 - (\varepsilon/(2t) - 2\tau)^2|}} \\
\times \begin{cases}
\mathbb{K}\left(\frac{(1 + \tau_{\rm a})^2 - (\varepsilon/(2t) + 2\tau)^2}{(1  - \tau_{\rm a})^2 - (\varepsilon/(2t) - 2\tau)^2}\right), & E_{\rm min} < \varepsilon < E_1, \\
\mathbb{K}\left(1 - \frac{(1 + \tau_{\rm a})^2 - (\varepsilon/(2t) + 2\tau)^2}{(1  - \tau_{\rm a})^2 - (\varepsilon/(2t) - 2\tau)^2}\right), & E_1 < \varepsilon <  E_2, \\
\mathbb{K}\left(\frac{(1 + \tau_{\rm a})^2 - (\varepsilon/(2t) + 2\tau)^2}{(1  - \tau_{\rm a})^2 - (\varepsilon/(2t) - 2\tau)^2}\right), & E_2 < \varepsilon < w_1;
\end{cases}
\end{multline}
\item $\tau_{\rm a}/2 < \tau < 1/2$: 
\begin{multline}\label{eq:rho_tau2}
\rho_{\rm R}(\varepsilon, \tau_{\rm a}, \tau) = \frac2{\pi^2t}\frac1{\sqrt{|(1 - \tau_{\rm a})^2 - (\varepsilon/(2t) - 2\tau)^2|}} \\
\times \begin{cases}
\mathbb{K}\left(\frac{(1 + \tau_{\rm a})^2 - (\varepsilon/(2t) + 2\tau)^2}{(1  - \tau_{\rm a})^2 - (\varepsilon/(2t) - 2\tau)^2}\right), & E_{\rm min} < \varepsilon < E_1, \\
\mathbb{K}\left(1 - \frac{(1 + \tau_{\rm a})^2 - (\varepsilon/(2t) + 2\tau)^2}{(1  - \tau_{\rm a})^2 - (\varepsilon/(2t) - 2\tau)^2}\right), & E_1 < \varepsilon <  w_1.
\end{cases}
\end{multline}
\begin{multline}
\rho_{\rm R}(\varepsilon, \tau_{\rm a}, \tau) = \frac2{\pi^2t}\frac1{\sqrt{|(1 + \tau_{\rm a})^2 - (\varepsilon/(2t) + 2\tau)^2|}} \\ \times\mathbb{K}\left(\frac{(1 - \tau_{\rm a})^2 - (\varepsilon/(2t) - 2\tau)^2}{(1 + \tau_{\rm a})^2 - (\varepsilon/(2t) + 2\tau)^2}\right), \\ w_1 < \varepsilon <  E_2.
\end{multline}
\item $1/2 < \tau$: 
\begin{equation}\label{eq:rho_tau3}
\rho_{\rm R}(\varepsilon, \tau_{\rm a}, \tau) = \frac2{\pi^2t} 
\begin{cases}
\frac{\mathbb{K}\left(\frac{(1 + \tau_{\rm a})^2 - (\varepsilon/(2t) + 2\tau)^2}{(1  - \tau_{\rm a})^2 - (\varepsilon/(2t) - 2\tau)^2}\right)}{\sqrt{|(1 - \tau_{\rm a})^2 - (\varepsilon/(2t) - 2\tau)^2|}}, & E_{\rm min} < \varepsilon< w_1, \\
2\frac{\mathbb{K}\left(\frac{(1 + \tau_{\rm a})^2 - (\varepsilon/(2t) + 2\tau)^2}{(1  - \tau_{\rm a})^2 - (\varepsilon/(2t) - 2\tau)^2}\right)}{\sqrt{|(1 - \tau_{\rm a})^2 - (\varepsilon/(2t) - 2\tau)^2|}}, & w_1 < \varepsilon <  w_2, \\
2\frac{\mathbb{K}\left(\frac{(1 - \tau_{\rm a})^2 - (\varepsilon/(2t) - 2\tau)^2}{(1  + \tau_{\rm a})^2 - (\varepsilon/(2t) + 2\tau)^2}\right)}{\sqrt{|(1 + \tau_{\rm a})^2 - (\varepsilon/(2t) + 2\tau)^2|}}, & w_2 < \varepsilon <  E_1, \\
\frac{\mathbb{K}\left(\frac{(1 - \tau_{\rm a})^2 - (\varepsilon/(2t) - 2\tau)^2}{(1  + \tau_{\rm a})^2 - (\varepsilon/(2t) + 2\tau)^2}\right)}{\sqrt{|(1 + \tau_{\rm a})^2 - (\varepsilon/(2t) + 2\tau)^2|}}, & E_1 < \varepsilon <  E_2,
\end{cases}
\end{equation}
\end{itemize}
where $\mathbb{K}(x)=\int_{0}^{\pi/2}(1-x\sin^2 \phi)^{-1/2}d\phi$ is the full elliptic integral of the first kind,
$E_{\rm min}  = 2t(-1 - \tau_{\rm a} - 2 \tau)$,
$E_1  = 2t(-1 + \tau_{\rm a} + 2 \tau)$, 
$E_2 = 2t(1 - \tau_{\rm a} + 2 \tau)$, 
$w_1 = 2t(1 + \tau_{\rm a} - 2 \tau)$,
$w_2  =  t\tau_{\rm a}/\tau$. 
%\begin{eqnarray}
%    E_{\rm min} & = & 2t(-1 - \tau_{\rm a} - 2 \tau),\\
%    E_1 & = & 2t(-1 + \tau_{\rm a} + 2 \tau), \\
%    E_2 & = & 2t(1 - \tau_{\rm a} + 2 \tau), \\
%    w_1 & = & 2t(1 + \tau_{\rm a} - 2 \tau), \\
%    w_2 & = &  t\tau_{\rm a}/\tau. 
%\end{eqnarray}

\begin{figure}[h!]
%\center{\includegraphics[scale=1,angle=270]{dos.pdf}\\}
\includegraphics[angle=-90,width=0.5\textwidth,clip]{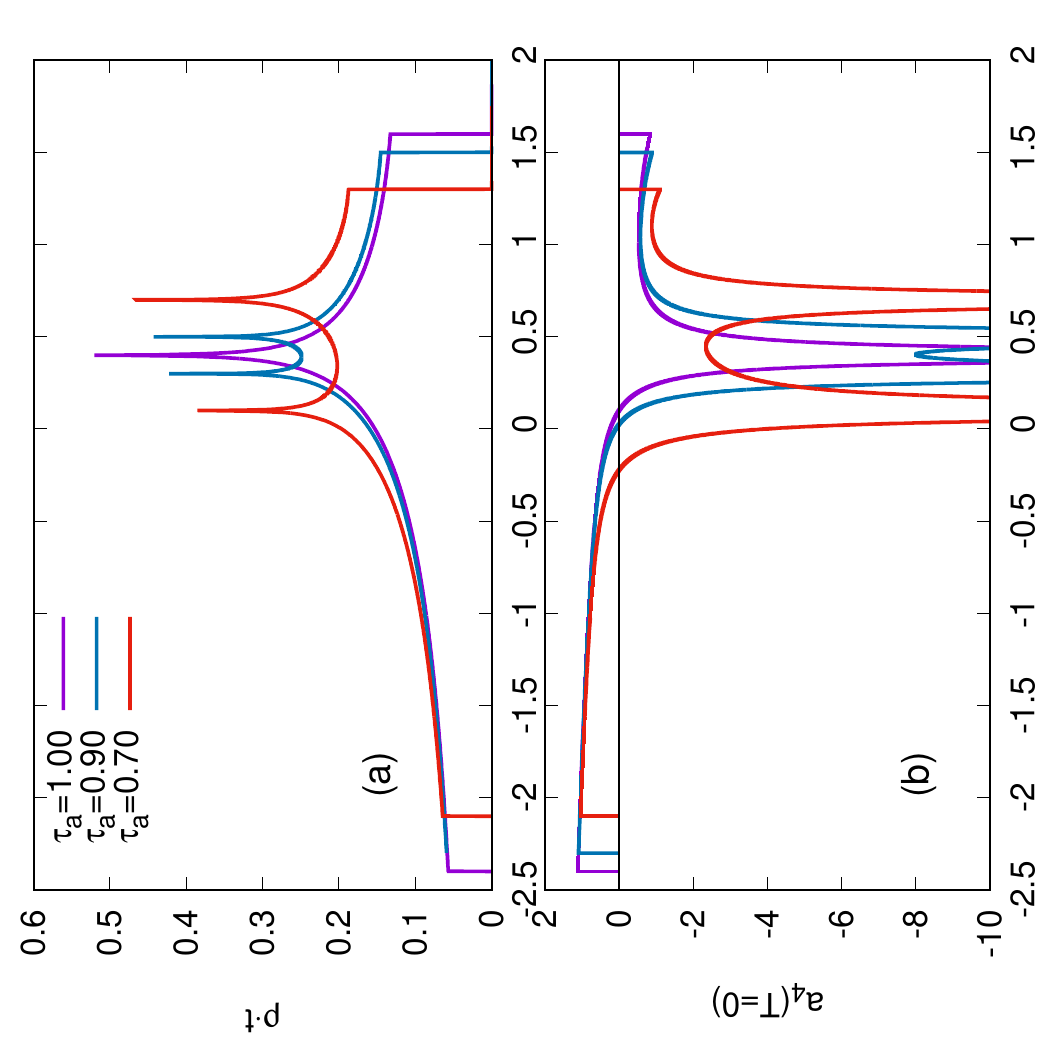}
\caption{
%{\green $[\times2]$} 
(a) Densities of states $\rho_{\rm R}(\epsilon, \tau_{\rm a}, \tau)$ of the rectangular lattice for different values of $\tau_{\rm a}$, at $\tau = 0.20$. (b) $a_4(T=0, \epsilon)$, see Eq.~(\ref{eq:a4_GS}). 
}
\label{fig:dos2d}
\end{figure}

At $\tau_{\rm a} = 1$, the spectrum of the square lattice without anisotropy with one van Hove peak of the density of states is realized, see Fig.~\ref{fig:dos2d}(a)~\cite{2017:Igoshev_MCE}. As $\tau_{\rm a}$ decreases, point symmetry of the lattice is broken  and this peak splits into two ones. A strong curvature of  DOS should be realized near and between the peaks, see Fig.~\ref{fig:dos2d};  corresponding energy dependence of the negative Landau expansion coefficient $a_4$ at $T = 0$ (see definition below in sec.~\ref{sec:Landau}) being also shown. 
In the context of IEMM transition, it is promising to set the PM phase Fermi level between the peaks. 

%\begin{figure}[h!]
%\center{\includegraphics[scale=1,angle=270]{dos.pdf}\\}
%\includegraphics[angle=-90,width=0.5\textwidth,clip]{d_dos_tau1=0.20.pdf}
%\caption
%{ {\green $a_4(T=0, \epsilon)$ (solid line, see Eq.~(\ref{eq:coefficients_ai:a4})) and $-1/(4\rho_{\rm R}(\epsilon, \tau_{\rm a}, \tau))^3 \times \rho_{\rm R}(\epsilon, \tau_{\rm a}, \tau)''/(3 \rho_{\rm R}(\epsilon, \tau_{\rm a}, \tau))$ (dashed line) of the rectangular lattice for different values of $\tau_{\rm a}$, at $\tau = 0.20$.  %Горизонтальная пунктирная линия показывает значение $\rho(\epsilon)$, обеспечивающее выполнение критерия Стонера $U\rho(\epsilon) = 1$ при $U = 1.5$t. 
%}}
%\label{fig:d_dos2d}
%\end{figure}

For $\tau<0.5$ at $\tau_{\rm a} = 1$, there are two van Hove saddle-type points in the spectrum at $\mathbf{k}=(0,\pi),(\pi,0)$, the corresponding
energy is $\epsilon=\epsilon_{\rm X} = 4\tau t$. These points produce a logarithmic van Hove singularity. When anisotropy is taken into account, the energies at the saddle points cease to coincide and the van Hove peak splits into two peaks with the energies $\epsilon_{\rm X,\pm}= 2t[2\tau + \pm(1 - \tau_{\rm a})]$. 
We consider two cases: 1) the strongly anisotropic case $\tau_{\rm a}=0.70$, $\tau=0.20$; 2) the weakly anisotropic case $\tau_{\rm a}=0.90$ and $\tau=0.20$, when $\tau_{\rm a}$ is close to unity. 

\begin{figure}[h!]
\includegraphics[angle=-90,width=\linewidth,clip]{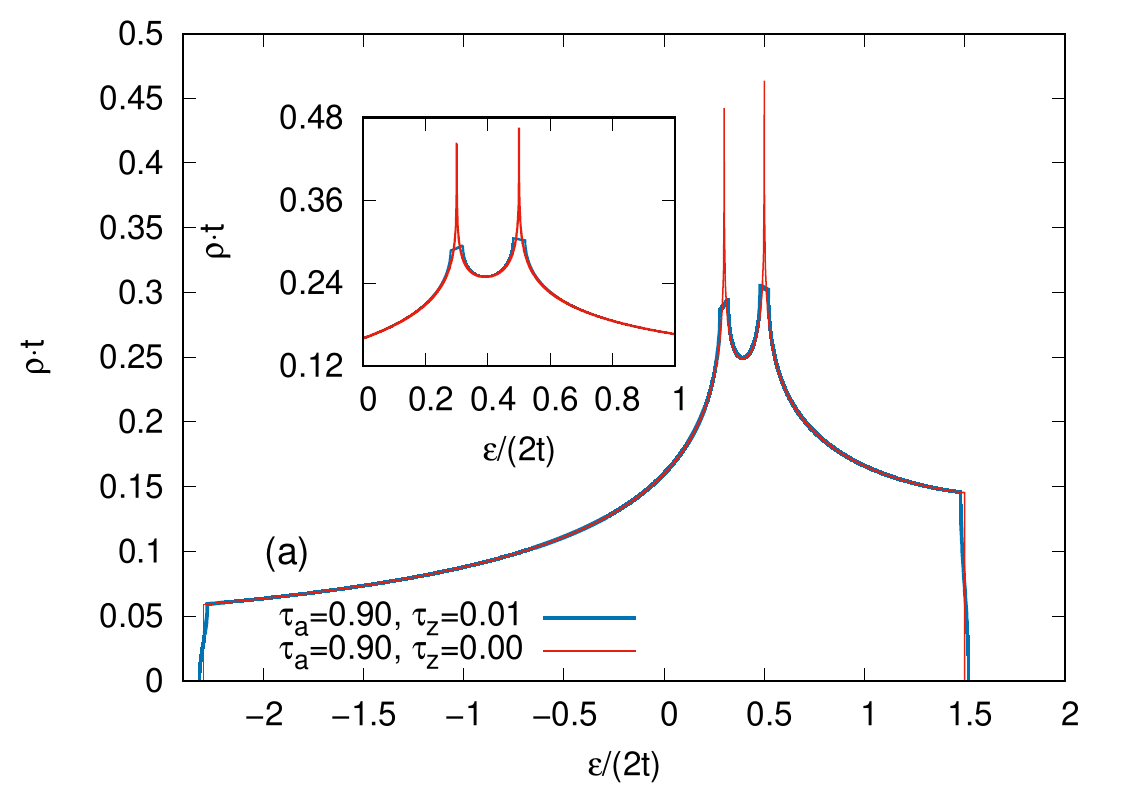}
\includegraphics[angle=-90,width=\linewidth,clip]{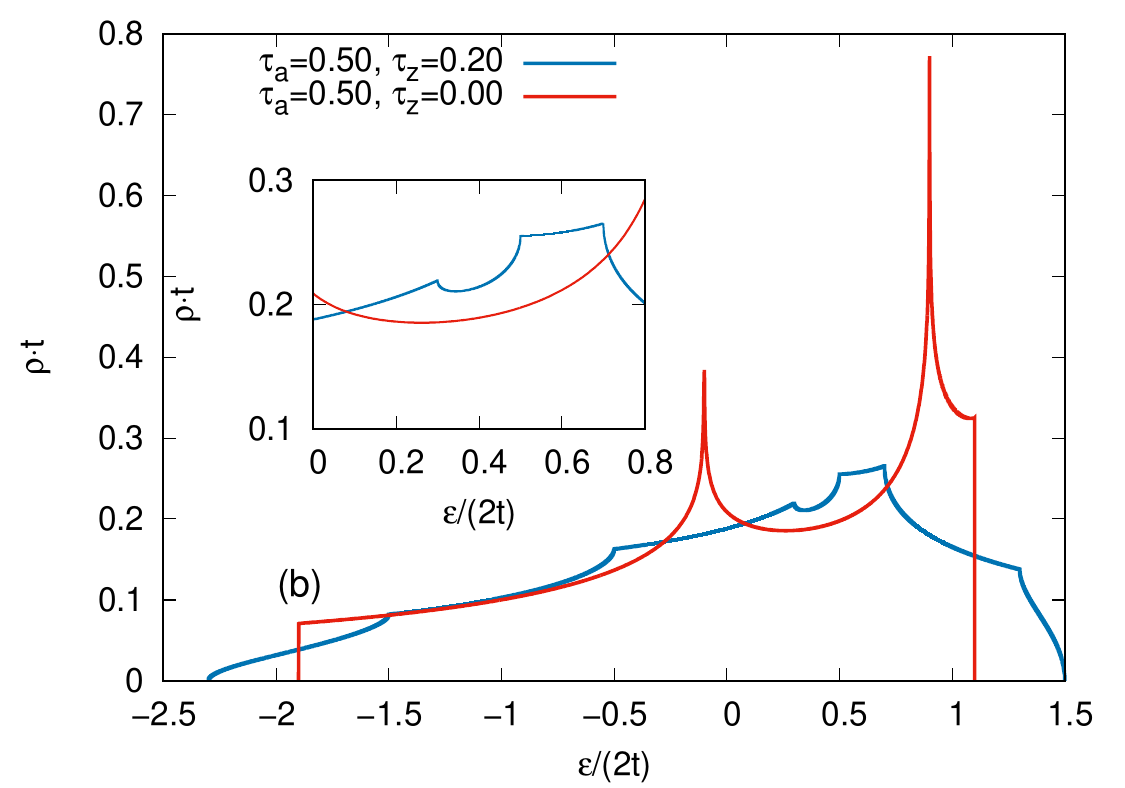}
\caption{
%{\green $[\times2]$} 
Density of states $\rho_{\rm OR}(\epsilon, \tau_{\rm a}, \tau, \tau_{\rm z})$ of  OR ($\tau_z \neq 0$) and rectangular $\rho_{\rm R}(\epsilon, \tau_{\rm a}, \tau)$ ($\tau_z=0$) lattice for $\tau = 0.20$. In the inset, the region of DOS near van Hove singularities is shown. 
(a)~$\tau_{\rm a} = 0.90,~\tau_z = 0.01$;  and 
(b)~$\tau_{\rm a} = 0.50,~\tau_z = 0.20$.
}
\label{fig:dos3d_an}
\end{figure}

The spectrum of a rectangular lattice can be extended to a three-dimensional case --- orthorhombic (OR) lattice %by adding a term $2 \tau_z \cos k_z$, $\tau_z$ 
including the hopping integral $t_z = t \tau_z$ along $z$ axis bond. Then, instead of one van Hove point, there appears a plateau (cf. \cite{2019:Igoshev:PMM:vHS,2019:Igoshev:JETP_Letters:vHS}), see~Fig.~\ref{fig:dos3d_an}. At that, the DOS $\rho_{\rm OR}$ between vHS levels preserves a strong curvature in two cases: (1) very small value $\tau_z$, we choose $\tau_z = 0.01$, see~Fig.~\ref{fig:dos3d_an}(a) and (2) sufficiently large value of $\tau_z$ at small value of $\tau_{\rm a}$, we choose $\tau_{\rm a} = 0.5$, $\tau_z = 0.2$, see~Fig.~\ref{fig:dos3d_an}(b). Increasing the hopping integral along the $z$ axis leads to plateau broadening and to large distortions of the rectangular lattice, so that preservation of the region with strong curvature is possible with increasing the energy distance between vHS positions (decreasing $\tau_{\rm a}$). 

%Taking into account only two transfer integrals is sufficient to model the giant van Hove singularity arising for the FCC lattice. 
The electron spectrum for FCC lattice in the nearest- and next-nearest-neighbor approximation %$\epsilon^{\rm FCC}_\mathbf{k}(\tau) = (1/N)\sum_{ij}t_{ij}\exp[\I\mathbf{k}(\mathbf{R}_i-\mathbf{R}_j)]$ 
has the following form  
\begin{multline}\label{eq:e(k)}
\epsilon^{\rm FCC}_\mathbf{k}(\tau)= \left[4\left(\cos\frac{k_x}2\cos\frac{k_y}2 + \cos\frac{k_x}2\cos\frac{k_z}2 \right.
\right.
\\
\left.
\left. + \cos\frac{k_y}2\cos\frac{k_z}2\right) - 2\tau(\cos k_x + \cos k_y + \cos k_z)\right]t.
\end{multline}

\begin{figure}[h!]
\includegraphics[angle=-90,width=0.5\textwidth,clip]{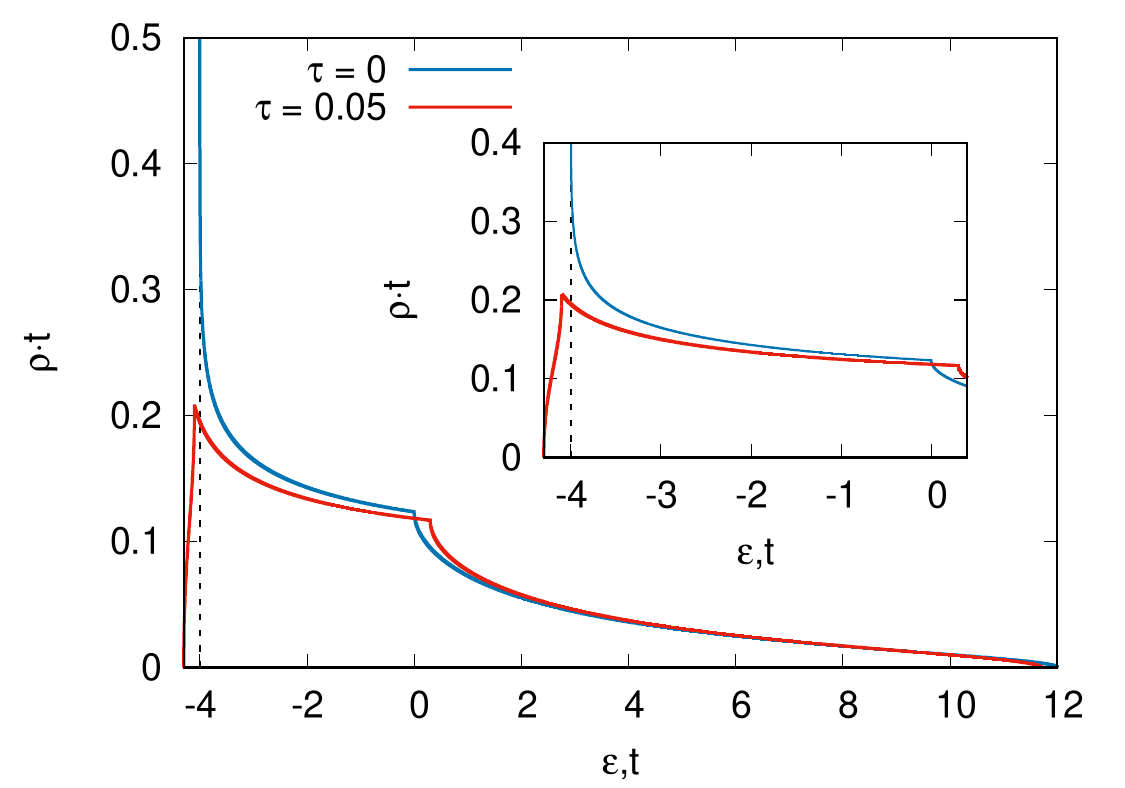}
\caption{The density of states $\rho_{\rm FCC}(\epsilon,\tau)$ for the FCC lattice for $\tau= 0$ and $0.05$. The inset shows the high-curvature region near the band bottom. 
}
\label{fig:dos_fcc_tau0}
\end{figure}

In the nearest-neighbor approximation, a logarithmic divergence in the dependence of $\rho_{\rm FCC}(\epsilon, \tau)$ at $\tau=0$  appears at the bottom of the band, see~Fig.~\ref{fig:dos_fcc_tau0}~\cite{1969:jelitto:density}. Near this feature, there is a region with strong positive DOS curvature, which can lead to  appearance of IEMM transition. 

In the case of small hopping between the next-nearest neighbors $\tau = 0.05$, the logarithmic divergence of DOS at the band bottom   becomes smeared and strong DOS curvature  is preserved, see~Fig.~\ref{fig:dos_fcc_tau0}. 

\begin{figure}[h!]
\includegraphics[angle=-90,width=0.5\textwidth,clip]{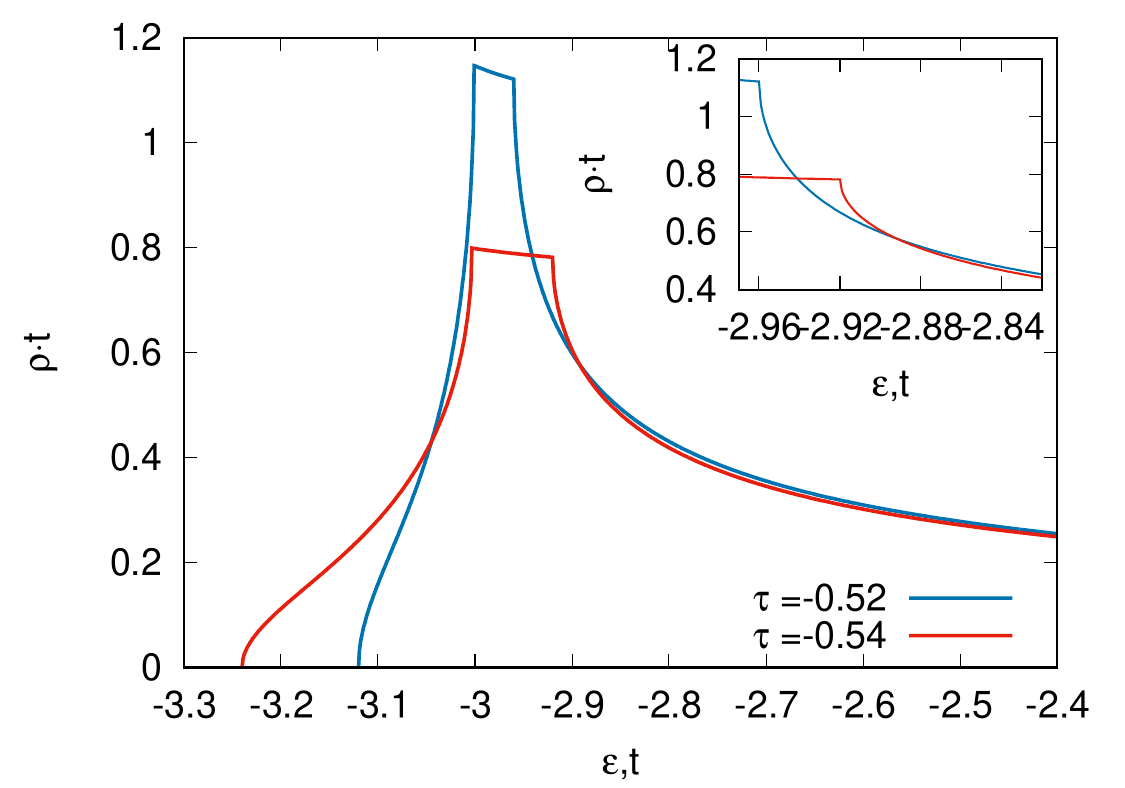}
\caption{The density of states $\rho_{\rm FCC}(\varepsilon,\tau)$ for the FCC lattice for $\tau=-0.52$ and $-0.54$. 
The inset shows the region of DOS where IEMM transition is considered, see below.
}
\label{fig:dos}
\end{figure}

At $\tau = -0.5$ we have giant (higher-order, quasi-one-dimension) vHS at the band bottom for FCC lattice with strong curvature~\cite{1998:Ulmke}. Such a case is ideal and  probably not realizable, so that we consider cases of $\tau$ which are close to $-0.5$. In the spectrum, the ratios of the transfer integral between the sites $\tau = -0.52$ and~$-0.54$ are taken into account, which ensures the appearance of a DOS feature  in the form of a van Hove plateau and a region with a strong DOS curvature, see~Fig.~\ref{fig:dos} and Ref.~\cite{2022:Igoshev:vHS_fcc}. Consideration of several values of $\tau$ will allow us to understand the impact of the DOS curvature  on the IEMM phenomenon. 

\section{Results}
In this section, the itinerant metamagnetic transition is investigated for the above-mentioned DOS's using  
numerical solution of Eqs. (\ref{eq:main_eqiations:n}) and (\ref{eq:main_eqiations:m}).  Equation~\eqref{eq:F_HFA} is used to determine the solution $m = m_i(T,n,h)$ with minimal free energy $F(T,n,h|m)$ for particular $T$, $n$ and $h$.  
%which the IEMM region and to investigate magnetic properties therein.

\subsection{Rectangular lattice}
In this section, we consider the IEMM transition for a rectangular lattice, see Sect.~\ref{sec:dos}. 
%was chosen due to the presence of a region between the peaks with strong curvature, so only cases with the Fermi level position in this region were considered.

\begin{figure}[h!]
\includegraphics[angle=270,width=\linewidth]{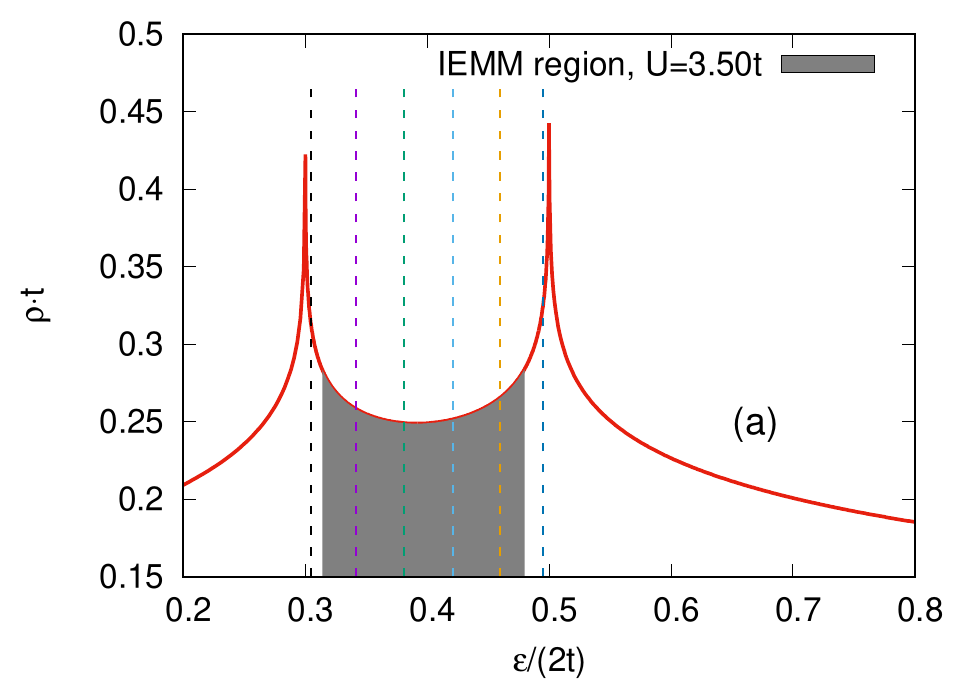}
\includegraphics[angle=270,width=\linewidth]{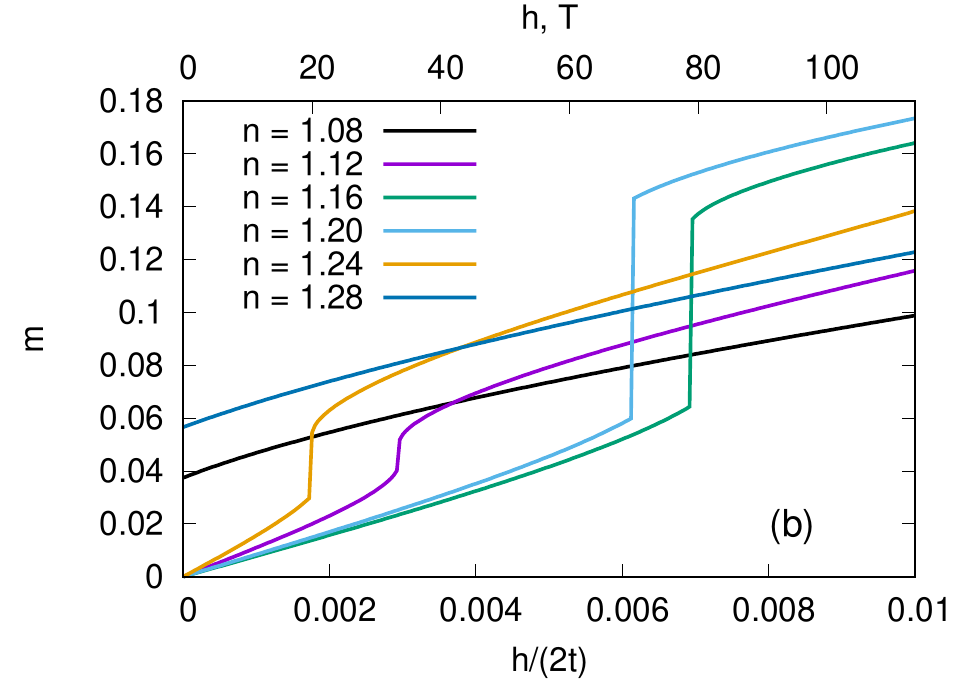}
\caption{%{\green $[\times2]$} 
(a) Density of states of a rectangular lattice $\rho_{\rm R}(\epsilon, \tau_{\rm a}, \tau)$ at $\tau_{\rm a}=0.90, \tau=0.20$. The vertical lines indicate PM phase Fermi level $E_{\rm F}$ corresponding to the chosen filling $n$. Filling shows the region in which the IEMM transition occurs at a given  $U=3.50 t$. (b) Magnetic field dependence of magnetization $m(h)$ for rectangular lattice $\tau_{\rm a}=0.90$, $\tau=0.20$ at $T=0.004 t$, $U=3.50 t$. The upper axis shows $h_{\rm c}$ in units of $T$, see text.
}
\label{fig:m_h_low_an}
\end{figure}

\begin{figure}[h!]
\includegraphics[angle=270,width=\linewidth]{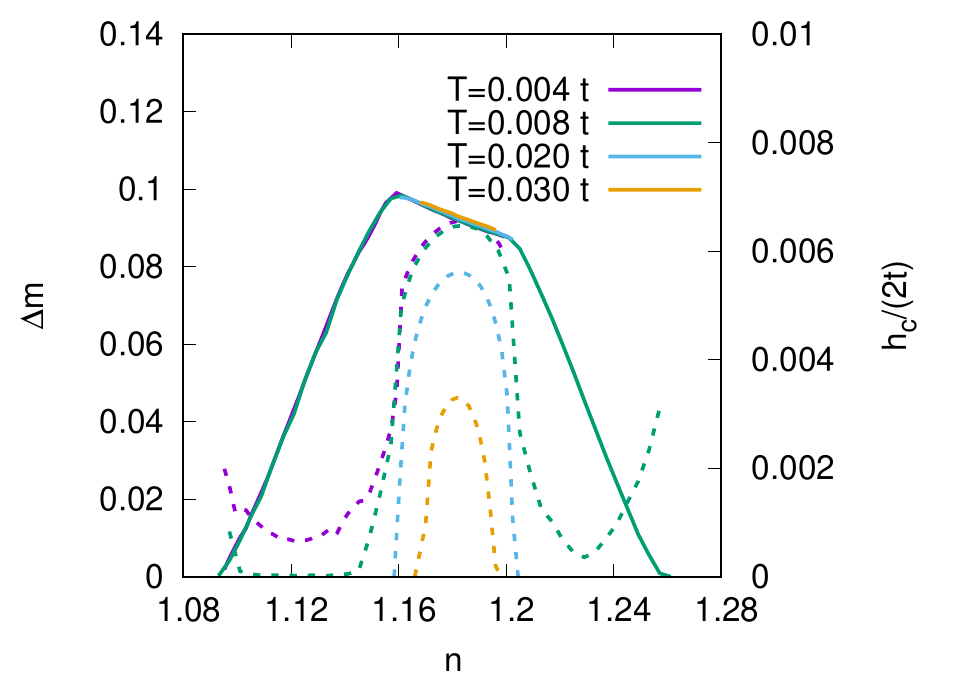}
\caption{%{\green $[\times2]$}
Dependences of magnetization jump $\Delta m$~(dashed line, left axis) and critical magnetic field $h_{\rm c}$~(solid line, right axis) on filling for rectangular lattice $\tau_{\rm a}=0.90$, $\tau=0.20$ at different temperatures at $U=3.50 t$. 
%The dashed line corresponds to $\Delta m (n)$, the solid line --- to $h_{\rm c}(n)$.
} 
\label{fig:gap_n_rec1}
\end{figure}

Choosing the filling where the Fermi level falls in the region between a pair of vHS levels, see Fig.~\ref{fig:m_h_low_an}(a),
we first study the magnetic field dependence $m = m_i(T,n,h)$  for the case of a weak anisotropy, $\tau_{\rm a}=0.90$, $\tau=0.20$ [see Fig.~\ref{fig:m_h_low_an}(b)], at low temperature $T=4\cdot10^{-3}t$ and $U=3.50 t$. 
This dependence is typical for IEMM transition: at  certain value of the external magnetic field, the magnetization exhibits a~magnetization jump $\Delta m$ indicating a first-order IEMM phase transition.
For convenience of comparison with typical experimental values, we choose the reference value of the bandwidth $W = W_{\rm R} = 7.6t$ to be 5~eV bandwidth. 
%To illustrate the results we choose $U = 7t$. 
%At the position of the Fermi level of PM phase at vHS positions, the magnetically ordered phase is realized, {\blue [Fedor, insert exact values: ]$n_{\rm vHS}=0.54,~0.64$}; as $E_{\rm F}$ moves towards the minimum $\rho_{\rm R}(\varepsilon)$ between vHSs, the IEMM transition appears with a small value of a~magnetization jump $\Delta m$ in relatively small fields $h_{\rm c}$ ($n=0.56,~0.62$){\blue[They should start from zero!]}. When the Fermi level is placed at the minimum of this DOS region, the phase transition occurs in large fields with a large magnetization jump. 

%\begin{figure}[h!]
%\includegraphics[angle=270,width=\linewidth]{m_h_tau2=0.2,n=0.56.pdf}
%\includegraphics[angle=270,width=\linewidth]{m_h_tau2=0.2,n=0.58.pdf}
%\caption{Dependence of magnetization on magnetic field $m(h)$ with temperature change for a rectangular lattice $\tau_{\rm a}=0.90$, $\tau=0.20$, $U=7.00 t$. (a) $n=0.56$; (b) $n=0.58$. 
%{\blue [Ne nujzno???]}}
%\label{fig:m_h_T}
%\end{figure}

\begin{figure}[h!]
\includegraphics[angle=270,width=\linewidth]{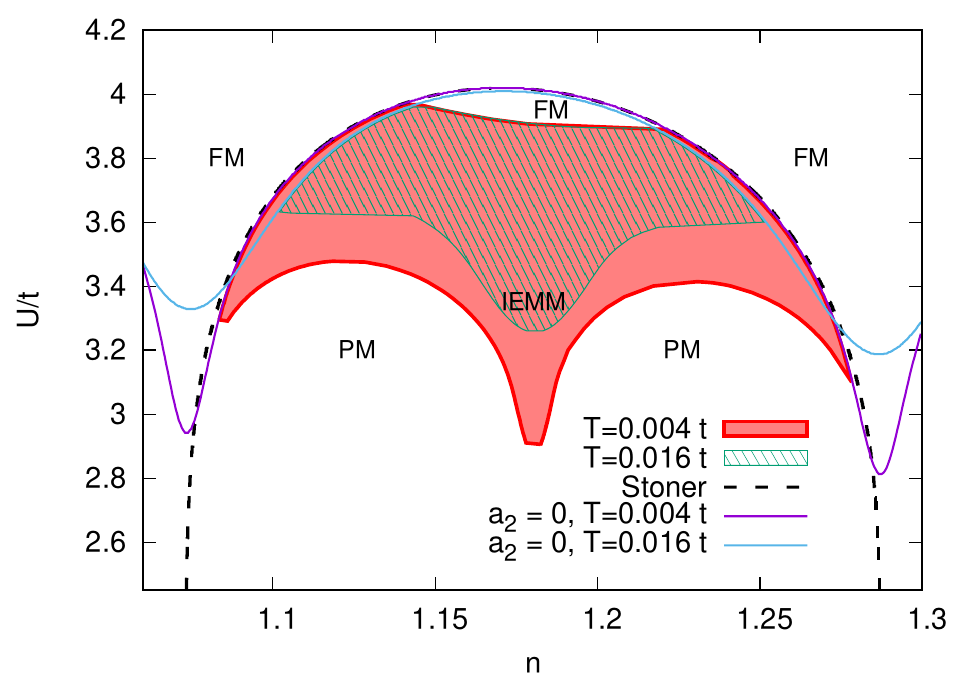}
\caption{%{\green $[\times2]$}
Phase diagram in $U-n$ plane for the rectangular lattice $\tau_{\rm a}=0.90$, $\tau=0.20$  at $T=0.004t$ and $0.016t$ with ferromagnetic (FM)  and paramagnetic (PM) phases,  and region of IEMM transition. The dashed line indicates the curve corresponding to the Stoner criterion $U \rho(E_{\rm F})=1$. Solid blue at $T = 0.016t$ and magenta at $T = 0.004t$ lines corresponds to $a_2 = 0$, see below section~\ref{sec:Landau}. 
}
\label{fig:f_d}
\end{figure}

%When the value of the electron filling $n$ changes, the dependence $m(h)$ changes significantly too. 
Important IEMM characteristics are the magnetization jump $\Delta m$ and the critical magnetic field $h_{\rm c}$ --- the magnetic field value at which $\Delta m$ occurs. 
The filling dependence of these characteristics at different temperatures are shown in~Fig.~\ref{fig:gap_n_rec1}. 
Whereas the  dependence $\Delta m(n)$ is strongly nonmonotonous, the $h_{\rm c}(n)$ dependence is dome-like and exhibits two kinks. 
Thus, at $U=3.50 t$ the IEMM transition region ($1.10 \lesssim n \lesssim 1.26$)  is inside the~FM~phase region. 
%When reaching the filling region where the IEMM transition occurs, ferromagnetism disappears, so that at filling corresponding to the transition boundary ($n\approx 0.55, 0.63$), we observe a non-zero magnetization jump at zero critical fields. Changing the filling leads to the PM phase Fermi level moving away from the peak of the density of states, which leads to a decrease in the magnetization jump and an increase in the critical field. 
When a PM phase Fermi energy approaches the minimum of $\rho(E_{\rm F})$, the $\Delta m$ and $h_{\rm c}$ are close to their maximum. 
With increasing temperature, the transition region narrows, the maximum value of the magnetization jump decreases, while the critical field filling dependence remains almost unchanged.
%Thus, the maximum magnetization jump is observed at filling values near the ferromagnetic phase and at $n$ corresponding to the position of the $E_{\rm F}$ between the DOS peaks in the minimum of $\rho(\varepsilon)$. 
%The value of the critical field increases as the Fermi level moves away from the DOS peaks. 

To characterize the $U$ dependence of 
IEMM transition, a phase diagram is calculated in the variables $U-n$ (see Fig.~\ref{fig:f_d}) at two temperature values $T = 4\times10^{-3}$ and $1.6\times10^{-2}t$.  The region bounded by red curves corresponds to IEMM region; the dashed line indicates the curve corresponding to the parameters $U$ and $n$, for which the Stoner criterion is fulfilled.

At the lower boundary of IEMM region, transition is not observed in the whole filling interval between the curves of fulfillment of the Stoner criterion, and  regions corresponding to PM phase  appear. In this case, IEMM transition occurs near the edge and the center of this filling interval. The latter is connected to the region with maximal DOS curvature.

In a real situation a competition with antiferromagnetic (AFM) N\'eel (or incommensurate) ordering should be considered. 
A detailed analysis of the magnetic phase diagram including the energy stability treatment in the Hartree-Fock and Kotliar-Ruckenstein slave-boson approximation was performed in Refs.~\cite{2010:Igoshev, 2013:Igoshev, 2015:Igoshev}. 
In particular, for the square lattice at $t' = 0.2t$ this treatment shows that the AFM state, being energetically stable at half-filling, is destroyed at doping and is changed by ferromagnetic and phase-separated magnetic states.
The electron filling interval corresponding to IEMM transition for rectangular lattice nearly corresponds to ferromagnetic transition region for the square lattice. 
One can also expect that breaking down the symmetry of $x$ and $y$ directions (passing from square to rectangular lattice) can suppress the AFM ordering due to splitting of vHS.

At the $U$ values from $3.50t$ to $3.80 t$, the filling interval of IEMM region is of maximum length. In this case, the IEMM transition is observed for any position of PM phase Fermi level between the peaks. Moving  away from IEMM region in filling forces first-order phase transition to FM phase. 

At larger values of $U$ (upper part of the diagram), the IEMM transition region is not bounded by the dashed line corresponding to the Stoner criterion. In the region between IEMM and FM phase regions, there is a FM region, in contrast to the Stoner criterion. 

The temperature increase gradually suppresses the IEMM transition region at lower $U$-boundary (see Fig.~\ref{fig:f_d}). At the upper $U$-boundary,  the IEMM transition is almost not affected by the temperature increase. 
The $a_0 = 0$ curves are strongly affected as temperature increases only at the edges of Stoner criterion's dome. 

%%%%%%%%%%%%%%%%%%%%%%%%%%%
% tau_a = 0.7
%%%%%%%%%%%%%%%%%%%%%%%%%%%

\begin{figure}[h!]
\includegraphics[angle=270,width=\linewidth]{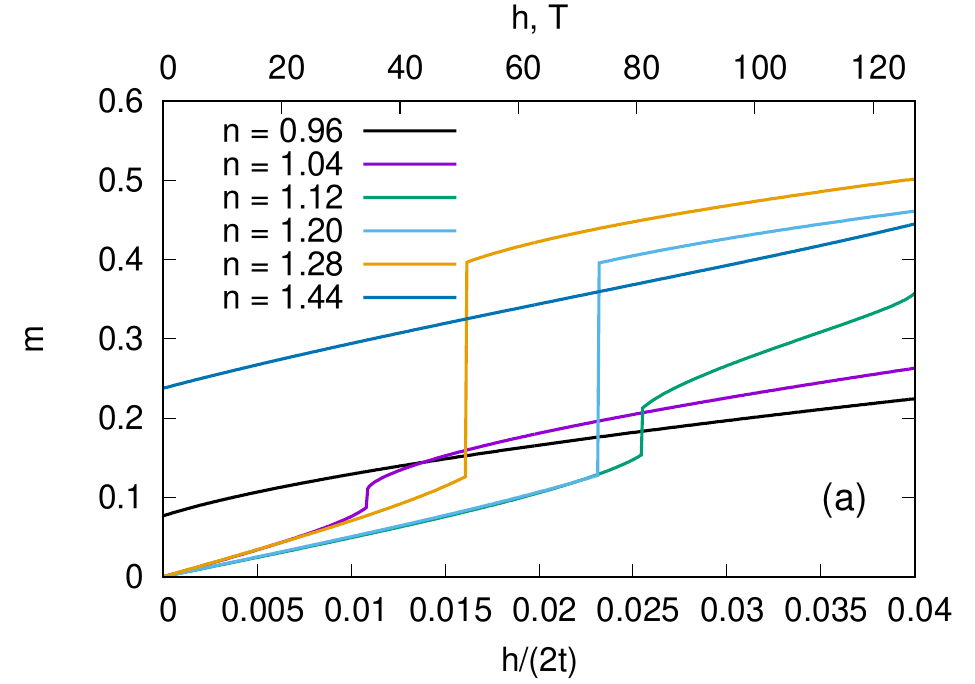}
\includegraphics[angle=270,width=\linewidth]{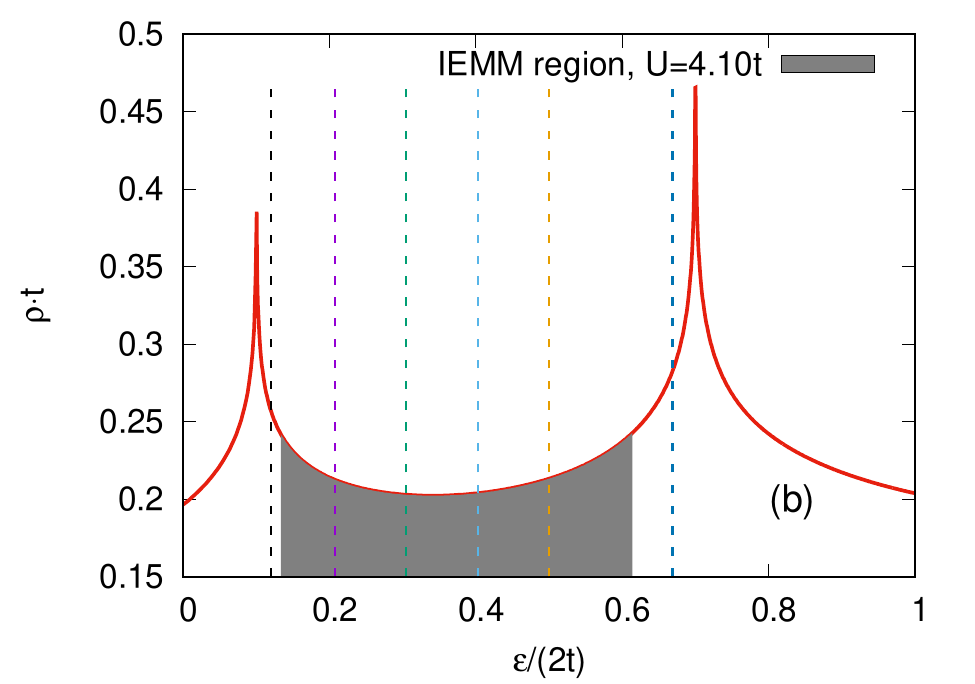}
\caption{%{\green $[\times2]$} 
(a) The same as for Fig.~\ref{fig:m_h_low_an} (b) for rectangular lattice $\tau_{\rm a}=0.70$, $\tau=0.20$ at $T=0.004 t$, $U=4.10 t$. (b) The same as for Fig.~\ref{fig:m_h_low_an} (a) for rectangular lattice $\rho_{\rm R}(\epsilon, \tau_{\rm a}, \tau)$ at $\tau=0.20$.
}
\label{fig:m_h_str_an}
\end{figure}

For the case of strongly anisotropic spectrum $\tau_{\rm a}=0.70$, $\tau=0.20$   at $T=0.004 t$, $U=4.10t$ the magnetic field dependence $m(T,n,h)$ has a  form, similar to the case of weakly anisotropic spectrum (see Fig.~\ref{fig:m_h_str_an}), but the IEMM transition occurs at larger $h_{\rm c}$ and $\Delta m$ reaches larger values. 
Then the filling interval corresponding to the IEMM transition increases. 
This is due to an increase in the energy difference between vHS's positions of $\rho_{\rm R}$ and corresponds to an increase of the region with strong DOS curvature.

\subsection{Orthorombic lattice} 

\begin{figure}[h!]
\includegraphics[angle=270,width=\linewidth]{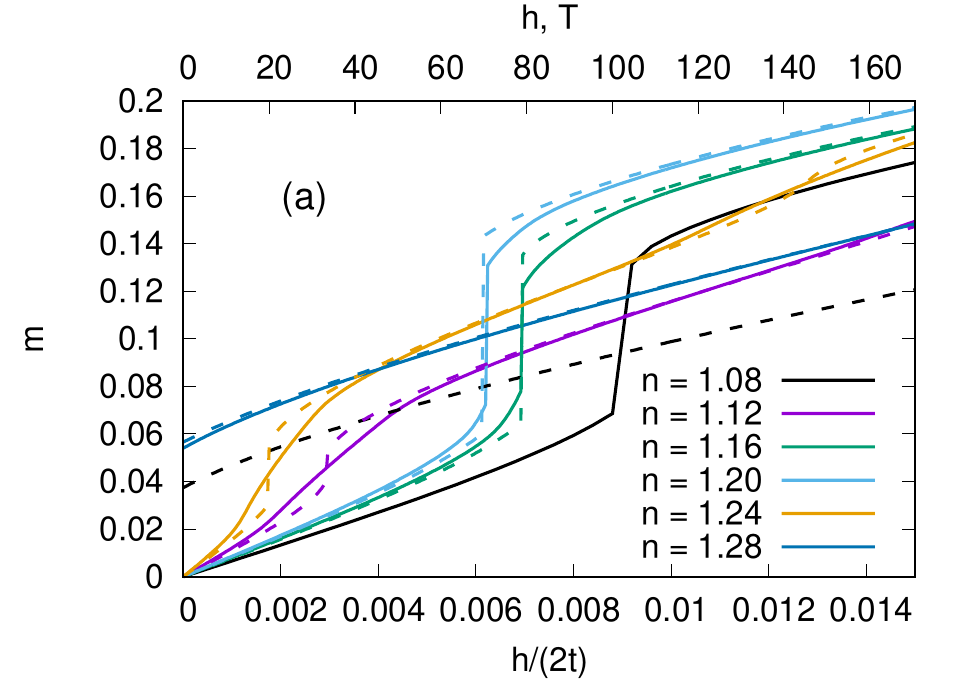}
\includegraphics[angle=270,width=\linewidth]{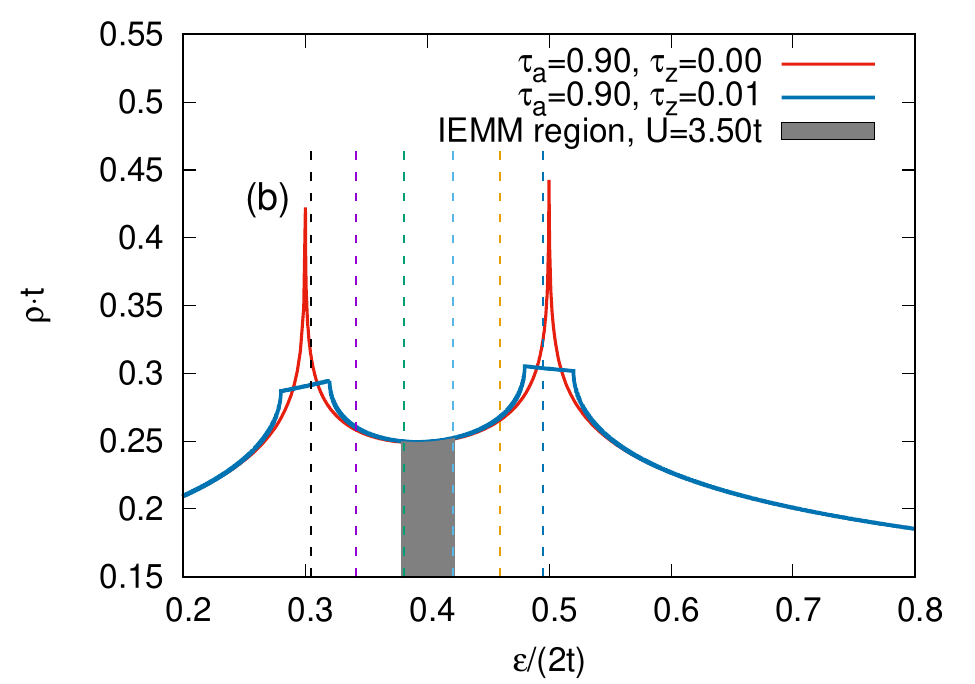}
\caption{%{\green $[\times2]$} 
The same as for Fig.~\ref{fig:m_h_low_an} (b) for OR lattice $\tau_z = 0.01$ (solid line) and the rectangular lattice (dashed line) $\tau_{\rm a}=0.90$, $\tau=0.20$ at $T=0.004 t$, $U=3.50 t$. (b) The same as for Fig.~\ref{fig:m_h_low_an} (a) for a rectangular and a OR lattice.
}
\label{fig:m_h_3d}
\end{figure}

\begin{figure}[h!]
\includegraphics[angle=270,width=\linewidth]{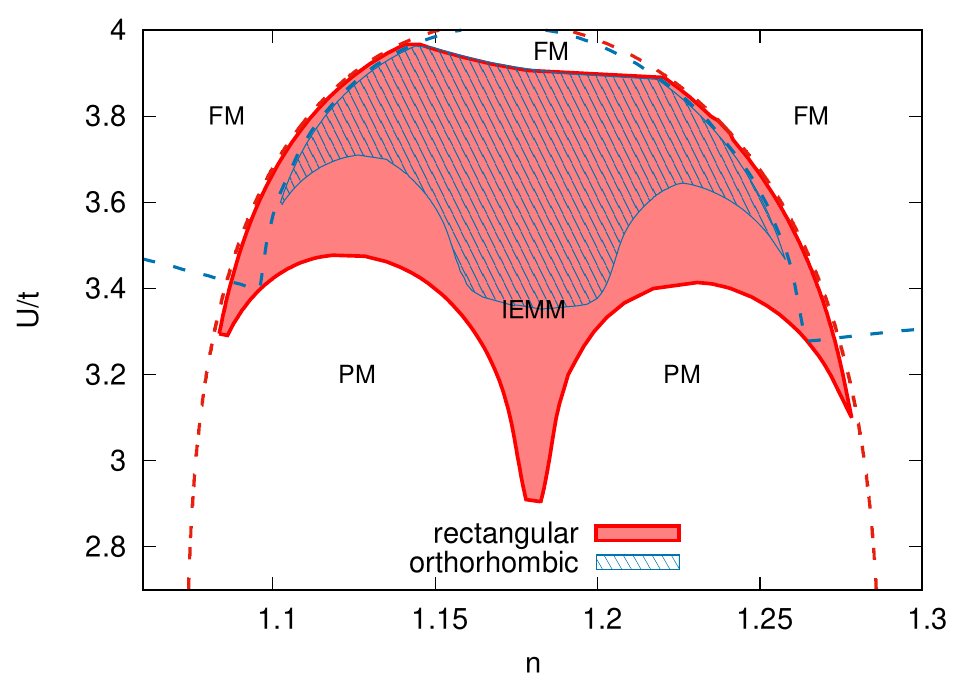}
\caption{%{\green $[\times2]$} 
Phase diagram in $U-n$ variables at $T=0.004t$ for rectangular and OR lattice $\tau_z = 0.01$, $\tau_{\rm a}=0.90$, $\tau=0.20$.
% with two phases: FM --- ferromagnetism, PM --- paramagnetism and region of metamagnetism --- MM. 
	The dashed lines indicates the curves corresponding to the Stoner criterion $U \rho(E_{\rm F})=1$ for both lattices.  
}
\label{fig:f_d2}
\end{figure}

\begin{figure}[h!]
\includegraphics[angle=270,width=\linewidth]{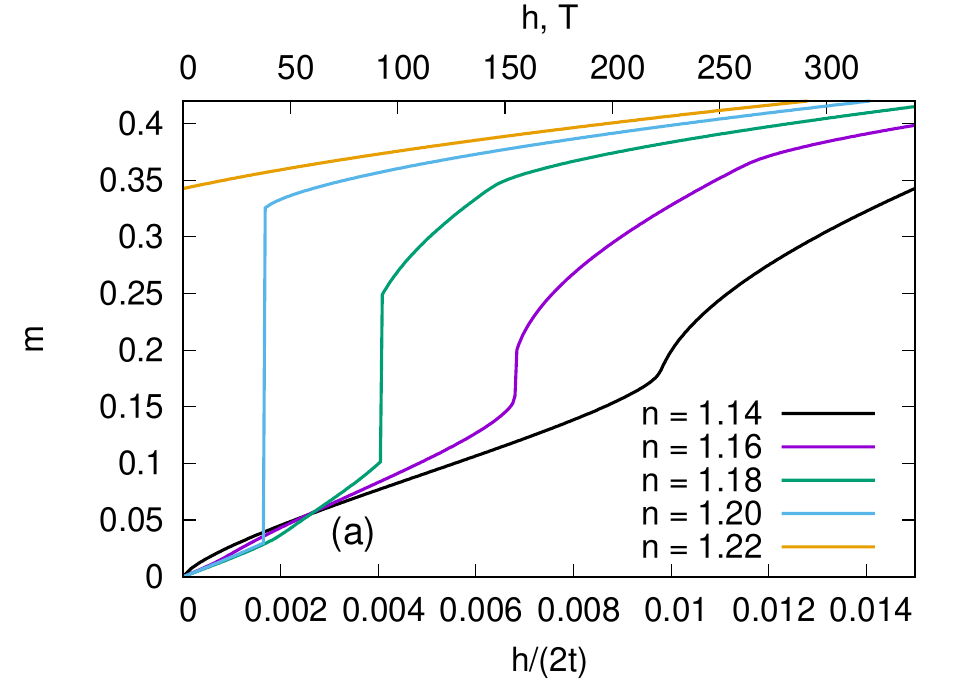}
\includegraphics[angle=270,width=\linewidth]{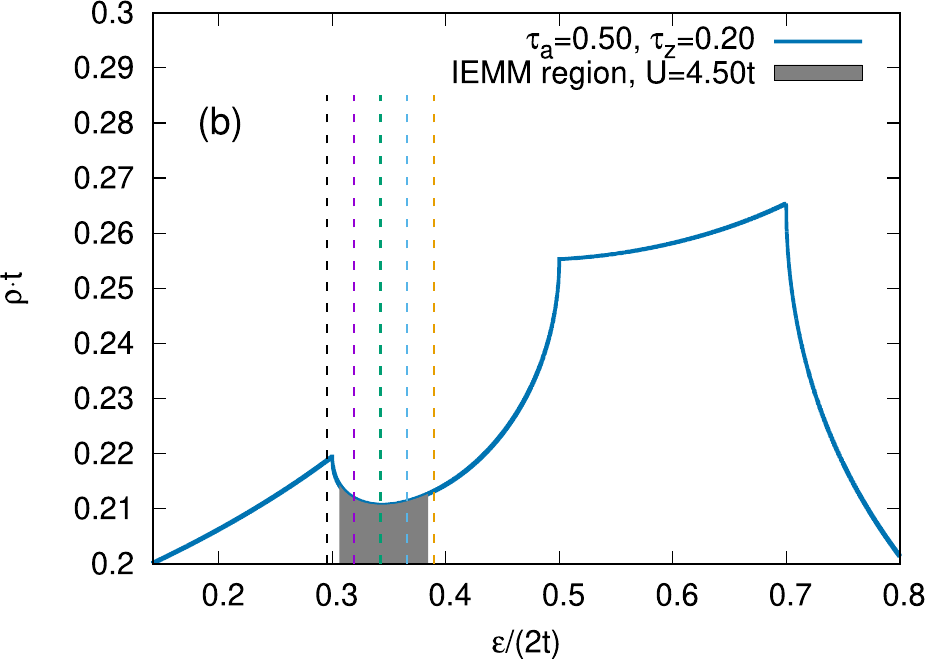}
\caption{%{\green $[\times2]$}
The same as for Fig.~\ref{fig:m_h_low_an} (b) for OR lattice $\tau_z = 0.20$, $\tau_{\rm a}=0.50$, $\tau=0.20$ at $T=0.004 t$, $U=4.50 t$. (b) The same as for Fig.~\ref{fig:m_h_low_an} (b) for OR lattice $\rho_{\rm R}(\epsilon, \tau_{\rm a}, \tau, \tau_{\rm z})$ at $\tau=0.20$. 
}
\label{fig:m_h_an_3d2}
\end{figure}

\begin{figure}[h!]
\includegraphics[angle=270,width=\linewidth]{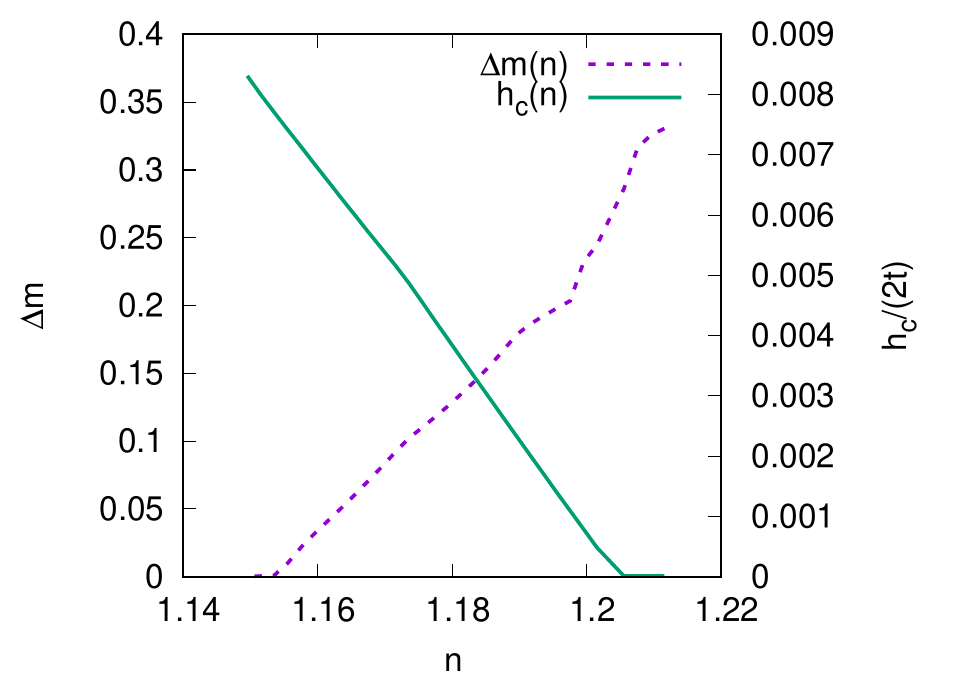}
\caption{%{\green $[\times2]$}
The same as for Fig.~\ref{fig:gap_n_rec1} for OR lattice $\tau_z = 0.20$, $\tau_{\rm a}=0.50$, $\tau=0.20$ at $T=0.004 t$, $U=4.50 t$.}
\label{fig:gap_n_ortor2}
\end{figure}

\begin{figure}[h!]
\includegraphics[angle=270,width=\linewidth]{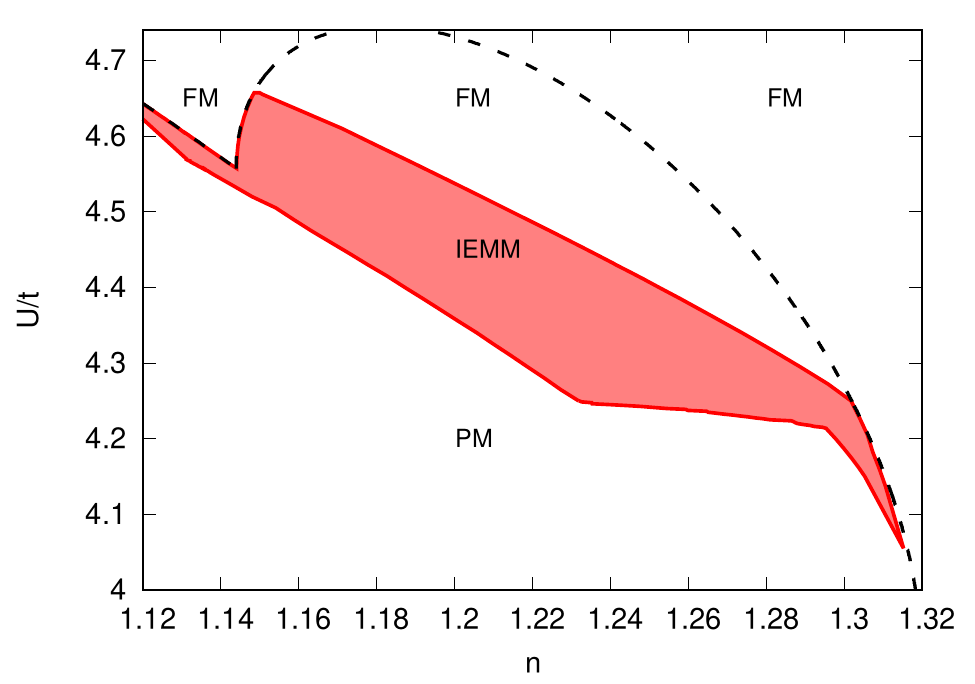}
\caption{%{\green $[\times2]$} 
Phase diagram in $U-n$ variables at $T=0.004t$ for OR lattice $\tau_z = 0.20$, $\tau_{\rm a}=0.50$, $\tau=0.20$
	% with two phases: FM --- ferromagnetism, PM --- paramagnetism and region of metamagnetism --- MM. 
The dashed line indicates the curve corresponding to the Stoner criterion $U \rho(E_{\rm F})=1$.  
The strong difference of FM condition from the Stoner one originates from the first order transition, see below in~the~Sec.~\ref{sec:Landau}.  
}
\label{fig:f_d3}
\end{figure}

%{\blue [Bad formulation]}\bad{As it has been already discussed, there are two possible cases considering the transfer along the $z$-axis} (more relevant case for real compounds), where the region with strong curvature of the DOS between vHS is preserved (see Fig.~\ref{fig:dos3d_an}).

A more realistic case is the spectrum including %taking into account 
the hopping along the $z$-axis, so that the rectangular lattice transforms into  orthorhombic one. For instance, we consider two cases in which the region with strong curvature of the density of states is preserved at $\tau = 0.2$: small $\tau_z = 0.01$ and $\tau_{\rm a} = 0.90$ [see~Fig.~\ref{fig:dos3d_an}(a)]; large $\tau_z = 0.20$ and $\tau_{\rm a} = 0.50$ [see~Fig.~\ref{fig:dos3d_an}(b)].

%%%%%%%%%%%%%%%%%%%%%%%%%%%
% small tau_z
%%%%%%%%%%%%%%%%%%%%%%%%%%%
%Let us first consider in detail the case in which the hopping along the $z$-axis is small, see DOS at~Fig.~\ref{fig:dos3d_an}(a).
In the case of the OR lattice with $\tau_z = 0.01$, $\tau_{\rm a}=0.90$, $\tau=0.20$, for the same parameters as considered above for the rectangular lattice ($U=3.50 t$, $T=0.004t$), magnetic field dependence of the magnetization is shown, see Fig.~\ref{fig:m_h_3d}(a). 
% (see Fig.~\ref{fig:m_h_3d}(b)), the~IEMM transition is retained. 
One can see that the IEMM transition for the OR lattice is partially suppressed at the edge of the IEMM filling interval for the rectangular lattice. 
%At the position of the Fermi level near singularities, where the densities of states of the two spectra differ, .
The comparison of the $U-n$ phase diagrams for rectangular  and OR lattices is presented in Fig.~\ref{fig:f_d2}. The replacement of DOS peaks by plateaus results in a decrease in the IEMM region, so that the transition is possible only at $U>3.3 t$. 

%%%%%%%%%%%%%%%%%%%%%%%%%%%
% large tau_z
%%%%%%%%%%%%%%%%%%%%%%%%%%%
We consider the case $\tau_z = 0.20$, $\tau_{\rm a} = 0.50$. Increasing the hopping along the $z$-axis causes the size of both plateaus to increase and the region with positive DOS curvature to decrease. At a certain value of $\tau_z$, the plateaus merge and the region with strong DOS curvature disappears. To preserve this region at larger values of $\tau_z$, it is necessary to increase the distance between the two plateaus. For this purpose we reduced the parameter $\tau_{\rm a}$, see Fig.~\ref{fig:m_h_3d}(b). 

In this case, the magnetic field  dependence of magnetization for  $U=4.50 t$ is presented in Fig.~\ref{fig:m_h_an_3d2}(a). The PM phase Fermi energy at the chosen filling values is in the region with positive curvature near the left plateau, see Fig.~\ref{fig:m_h_an_3d2}(b). The  increase of filling results in  monotonous increasing the magnetization jump $\Delta m$   and decreasing the critical field $h_{\rm c}$  (see~Fig.~\ref{fig:gap_n_ortor2}).

For this case, the phase diagram in $U-n$ variables is shown in~Fig.~\ref{fig:f_d3}. 
The kink of IEMM transition line at $n = 1.15$ is caused by crossing of the PM Fermi level and the vHS energy level, see~Fig~\ref{fig:m_h_an_3d2}(b). 
This situation substantially differs from that considered above for a rectangular lattice (cf.~Fig.~\ref{fig:f_d}). Decreasing the parameter $U$ leads to a shift of the IEMM transition region towards larger $n$, which does not agree with the Stoner criterion curve behavior. 

%\bad{
%This is due to the fact that the IEMM transition at smaller values of $U$ requires a larger value of DOS near the $E_{\rm F}$, and when $n$ increases, the Fermi level in the paramagnetic phase shifts to a region with a larger density of states. 
%}

\subsection{FCC lattice}
In this section, we treat IEMM transition in the 3D FCC lattice for the cases of small $\tau$ and $\tau \simeq -0.5$. Both these cases correspond to strong vHS. We choose the reference value of the bandwidth $W = 16 t$ for  $\tau=0, 0.05$ and $W=18 t$ for  $\tau \simeq -0.5$ to be 5~eV.

\subsubsection{$\tau = 0$}

The dependence of the density of states for the FCC lattice spectrum in the nearest-neighbor hopping approximation ($\tau = 0$) has a logarithmic divergence at the band bottom, see~Fig.~\ref{fig:dos_fcc_tau0}(a) in Sec.~\ref{sec:dos}. As a consequence, a region with positive curvature $\rho(\epsilon)$ appears at the bottom of the band, which, as in the case of an orthorhombic lattice, can lead to IEMM transition when PM phase Fermi level is positioned in high-curvature region between two vHS levels. 
%\sout{Let us consider how in this case the character of the transition will change, for this purpose we analyze the field and temperature dependences, as well as the phase diagram in $U-n$ parameters.}

\begin{figure}[h!]
\includegraphics[angle=270,width=\linewidth]{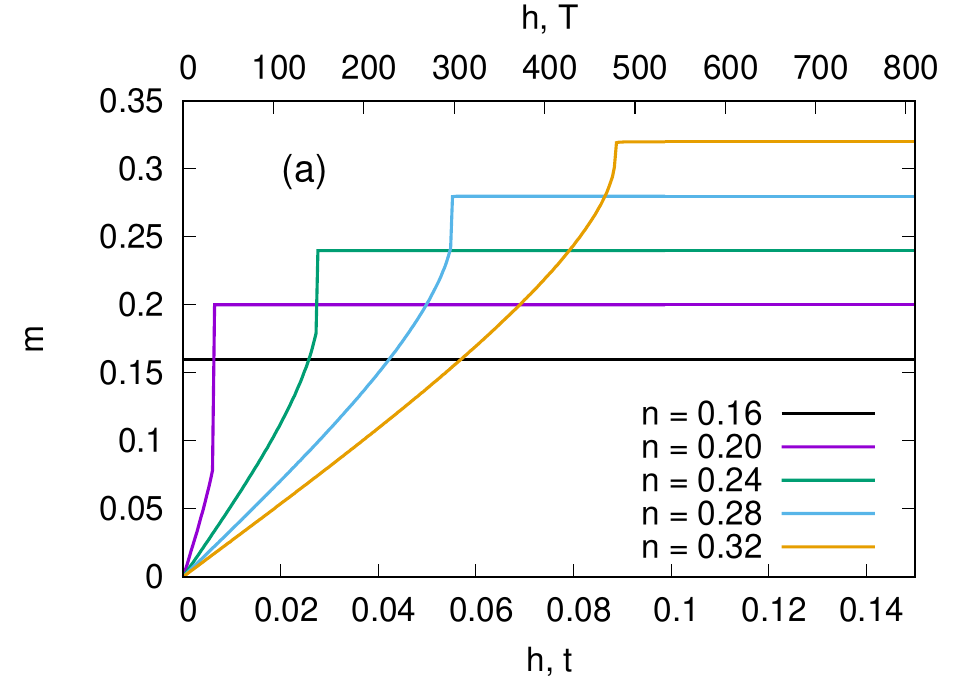}
\includegraphics[angle=270,width=\linewidth]{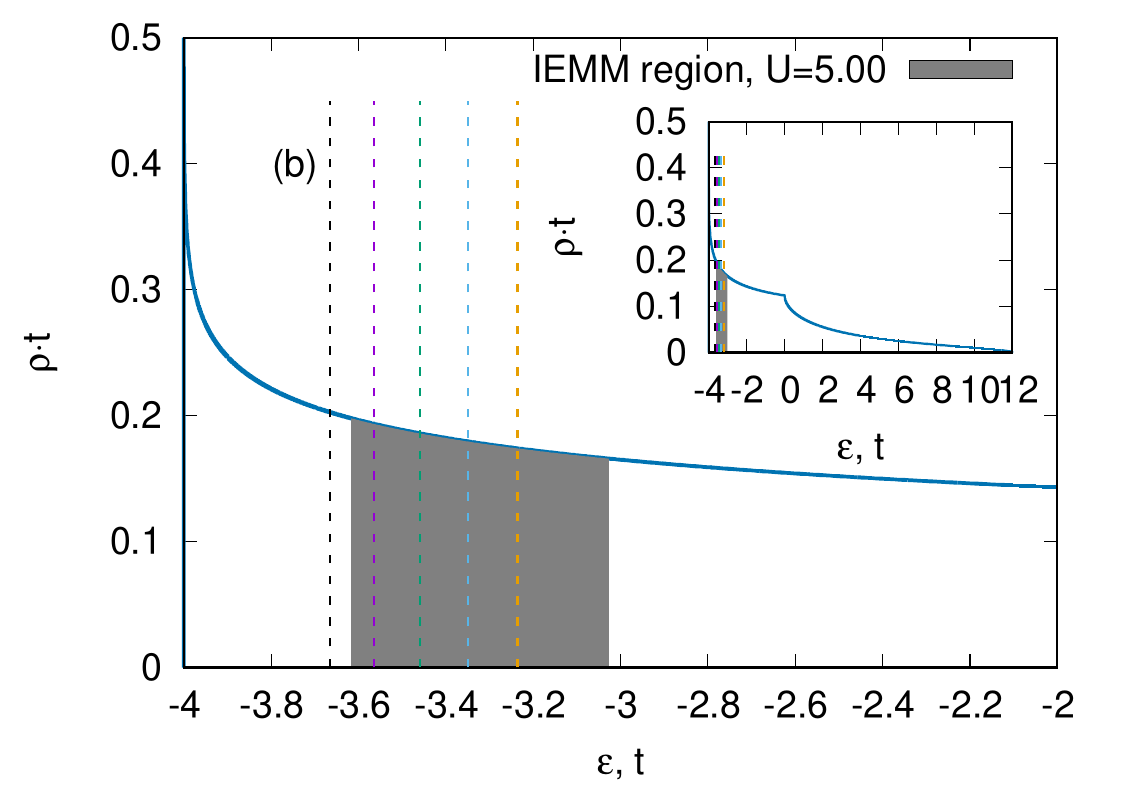}
\caption{(a) The same as for Fig.~\ref{fig:m_h_low_an}(b) for FCC lattice $\tau=0$, at $T=0.002 t$, $U=5.00 t$. (b)  The same as for Fig.~\ref{fig:m_h_low_an} (a) for FCC lattice $\rho_{\rm FCC}(\varepsilon,\tau=0)$.
}
\label{fig:m_h_fcc_tau0}
\end{figure}

\begin{figure}[h!]
\includegraphics[angle=270,width=\linewidth]{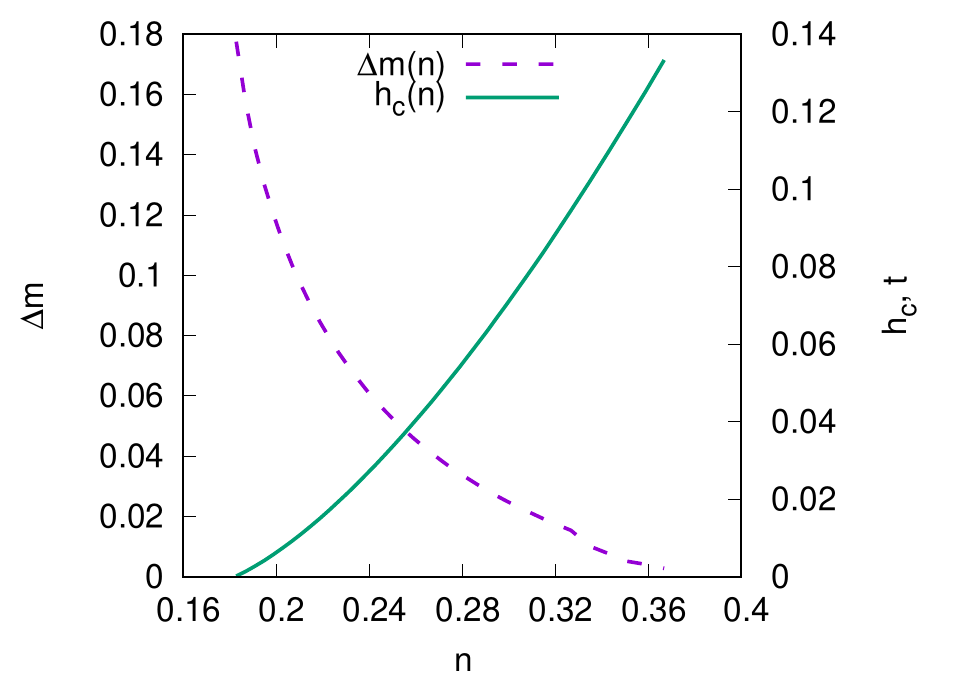}
\caption{The same as for Fig.~\ref{fig:gap_n_rec1} for FCC lattice at $\tau=0$,  $T=0.002 t$, $U=5.00 t$. 
}
\label{fig:gap_n_fcc1}
\end{figure}

%\bad{The presented magnetic field dependence $m(h)$ exhibit the magnetization jump (see~Fig.~\ref{fig:m_h_fcc_tau0}) as for the lattices considered above. An increase in filling leads to a decrease in the density of states at the Fermi level, resulting in an increase in the critical transition magnetic field $h_{\rm c}$ and a decrease in the magnetization jump. The behavior of the dependences is similar to those presented in \cite{levitin1988itinerant} for the schematic density of states, where the existence of metamagnetism is also associated with a special behavior of the density of states near the Fermi energy.} 
The magnetic field magnetization dependence for this case is shown in Fig.~\ref{fig:m_h_fcc_tau0}(a) and the positions of PM phase Fermi level on DOS plot corresponding to different filling values are shown in~Fig.~\ref{fig:m_h_fcc_tau0}(b). 
One can see that the IEMM transition to saturated FM is observed.
A filling increase leads to a monotonous decrease in $\Delta m$ and an increase in $h_{\rm c}$ (see~Fig.~\ref{fig:gap_n_fcc1}). 
While DOS plot seems to be different from that for the rectangular lattice, some similarity (large critical magnetic field values) of these two cases is due to the fact that there is strong curvature region between two vHS peculiarities of DOS in both the cases. 

%With increasing temperature, the magnetization jump  gradually decreases and the transition to saturation of magnetization  becomes smoother, see~Fig.~\ref{fig:m_T_fcc_tau0}. At $n=0.20$, the jump of magnetization is preserved at higher temperatures than at $n=0.28$, see~Fig.~\ref{fig:m_T_fcc_tau0}. 

%\begin{figure}[h!]
%\includegraphics[angle=270,width=\linewidth]{m_T_fcc_tau01.pdf}
%\caption{Magnetic field dependence of magnetization  $m(h)$ for different temperatures (shown in the legend) in the FCC lattice with $\tau = 0$, $U=5.00t$, $n = 0.20$ and $n = 0.28$. The upper axis is given in units of T.
%}
%\label{fig:m_T_fcc_tau0}
%\end{figure}

\begin{figure}[h!]
\includegraphics[angle=270,width=\linewidth]{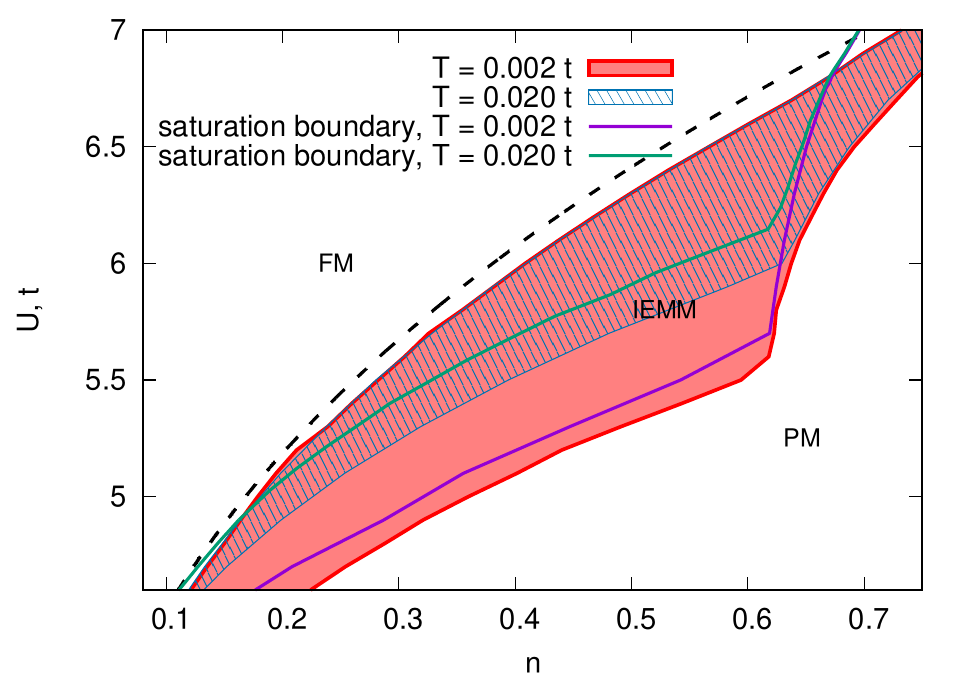}
\caption{Phase diagram in $U-n$ variables for FCC lattice for $\tau = 0$ at different temperatures: FM is ferromagnetic region, PM is paramagnetic region. Within the latter, filled IEMM areas correspond to metamagnetism regions. The dotted line indicates the curve corresponding to the Stoner criterion $U \rho(E_{\rm F})=1$, the saturation boundary line divides the region of saturated and unsaturated magnetism at $T=0.002 t$ and $0.020 t$. 
%{\blue [Why IEMM transition does not depend on temperature value at large $U$?]} 
%{\red 
%MM oblst' ochen shiroka, nuzhno kak-to pojasnit' eto? %- pri bolshih n skachok momenta mal i krit. pole h ochen veliko?
%Verojatno, pri tau=0 nuznhno brat' $U<5$, chtoby poluchit' fizicheski razumnye h.
% Pri tau=0.05  razumnye h - pri U=5.5. 
%Malye vs bolshie n?
%Mozhno takzhe pokazat' vtoruju critical line, kotoraja sootvetstvuet oblasti saturated FM (net paraprocessa) i ttetju, kotoraja sootvetstvuet oblasti saturated MM (skachok do maksimalno vozmozhnogo momenta).
%Vozmozhno, pri malyh tau eti linii budut sovpadat', no ne tak pri tau = -0.52 i dlya OR reshetki.
%}
}
\label{fig:f_d_fcc1}
\end{figure}

%\sout{The field dependences show that the behavior of the metamagnetic phase transition is strongly filling dependent. 
%In order to find out in detail the influence of parameters $n$ and $U$ on metamagnetism,} 
The $U-n$ phase diagram at $T=0.002t$ is presented in Fig.~\ref{fig:f_d_fcc1}. 
%\sout{Due to the absence of a two-peak DOS structure, in this case there is no situation in which the metamagnetic region would be between two ferromagnetic phases. }  
In this case, the broad IEMM region located between the two vHS levels separates  two phases: PM from below and FM from above. The saturation boundary, being in the region of IEMM transition, separates the regions of saturated and unsaturated magnetism in critical fields. Almost in all regions of metamagnetic transition, there is saturated magnetism in $h = h_{\rm c}$, only at $U>6.80t$ and $n>0.67$ the saturation boundary goes to the FM region. In the FM region, the saturation boundary already separates saturated and unsaturated ferromagnetism.

There are large $\Delta m$ and $h_{\rm c}$ values in this case, which is related to large energy interval of high DOS curvature. 
%This relation is similar to the case FCC lattice with $\tau = 0$. 
The same anomalously large critical magnetic fields of IEMM transition, and even larger ones are observed for a square lattice with $\tau = 0.35$ \cite{yamase2023ferromagnetic}, where IEMM transition is also caused by the strong curvature of DOS between the vHS peak and the vHS at the bottom or top of the band.
%{\blue [Fedor, please, make data for Fermi level of spin subband as a function of $n$ and $h$.] }
%{\red[Also present a DOS picture vs. magnetic field demonstrating crossing the Fermi level by the DOS peak for one spin projection to demonstrate saturated FM and large jump.]}

\subsubsection{$\tau = 0.05 $}
\begin{figure}[h!]
\includegraphics[angle=270,width=\linewidth]{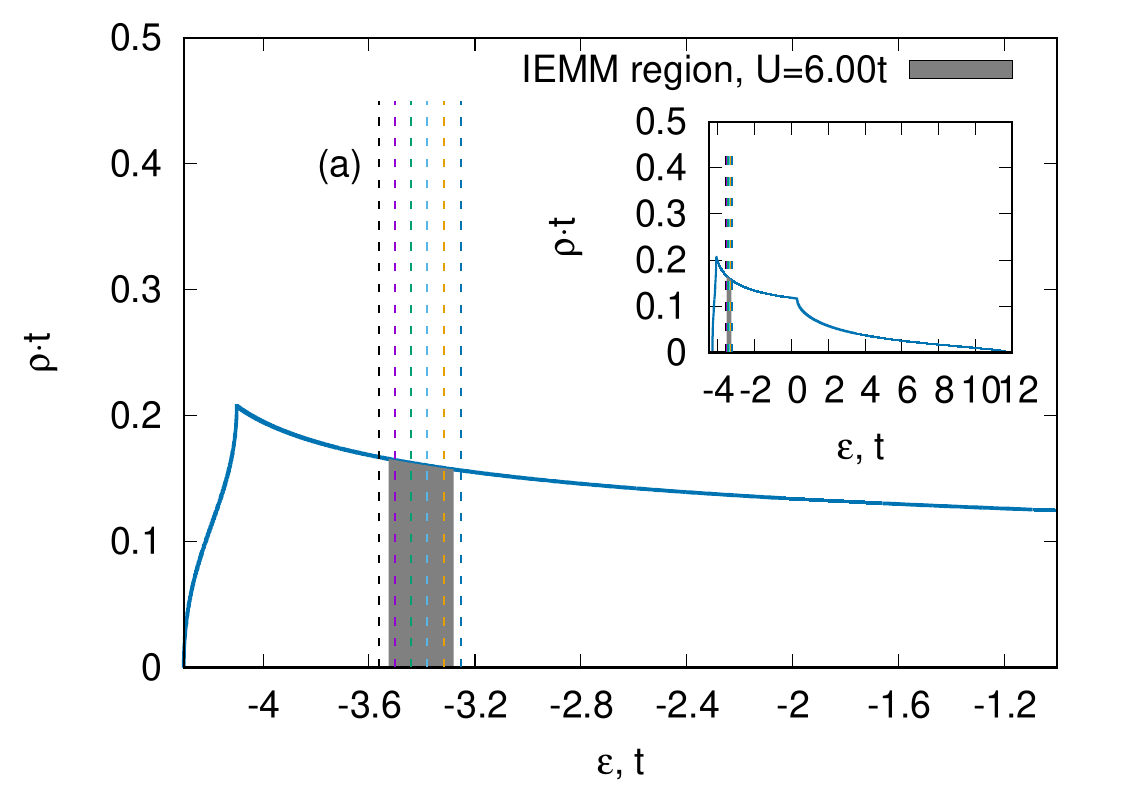}
\includegraphics[angle=270,width=\linewidth]{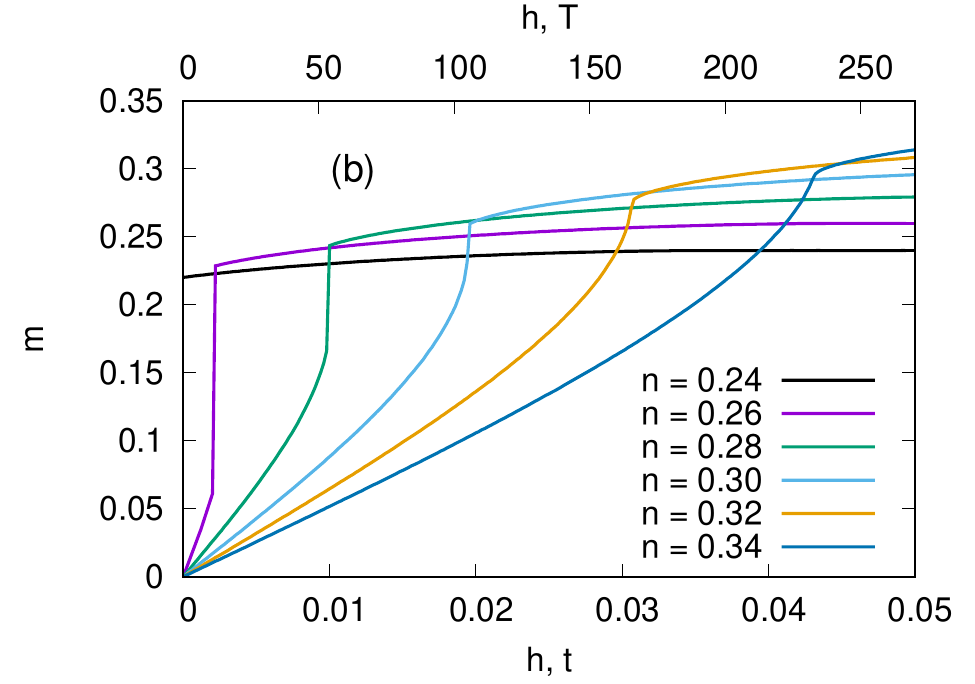}
\caption{(a) The same as for Fig.~\ref{fig:m_h_low_an} (b) for FCC lattice $\tau=0.05$, at $T=0.002 t$, $U=6.00 t$. (b)  The same as for Fig.~\ref{fig:m_h_low_an} (a) for FCC lattice $\rho_{\rm FCC}(\varepsilon)$ $\tau=0.05$.  
}
\label{fig:m_h_fcc_tau0.05}
\end{figure}

A more realistic case of the spectrum for the FCC lattice is that taking into account the next-nearest hopping integral. We assume it to be small, $\tau = 0.05$. Then the divergence at the bottom of the band observed at $\tau = 0$ is replaced by a plateau in the vicinity of band bottom, see~Fig.~\ref{fig:m_h_fcc_tau0.05}(a). 

\begin{figure}[h!]
\includegraphics[angle=270,width=\linewidth]{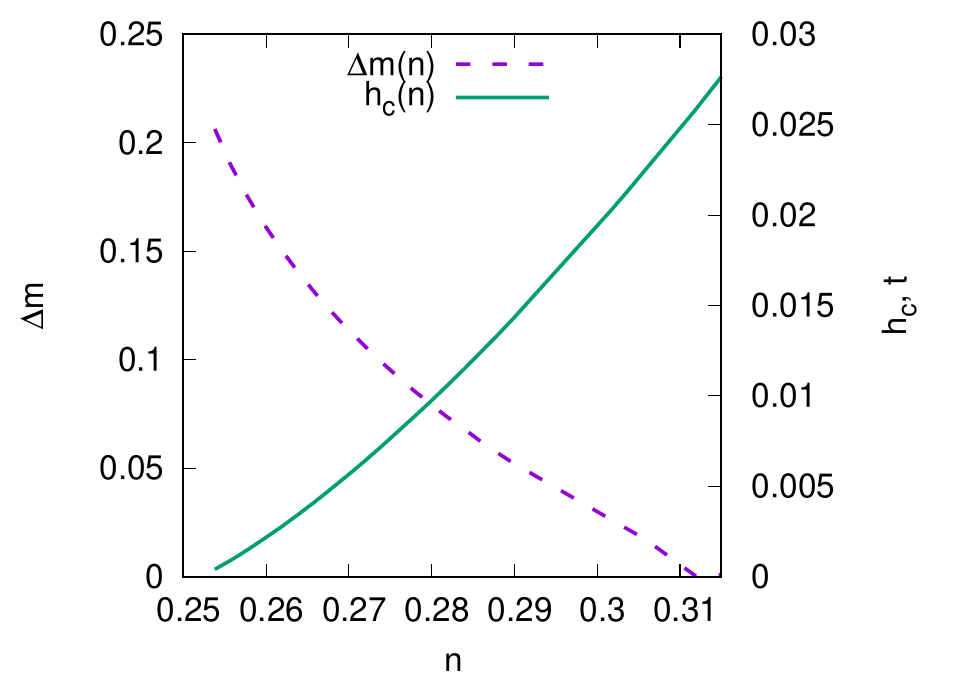}
\caption{The same as for Fig.~\ref{fig:gap_n_rec1} FCC lattice with $\tau=0.05$, at $T=0.002 t$, $U=6.00 t$. 
}
\label{fig:gap_n_fcc_tau2}
\end{figure}

For such DOS, the magnetic field dependences at the IEMM transition, see~Fig.~\ref{fig:m_h_fcc_tau0.05}, exhibits remarkable \textit{paraprocess}  (an increase of the spontaneous magnetization of a ferromagnet in the ordered phase under an external magnetic field) at $h > h_{\rm c}$, in contrast to the case $\tau=0$ (see~Fig.~\ref{fig:m_h_fcc_tau0}(a)).

\begin{figure}[h!]
\includegraphics[angle=270,width=\linewidth]{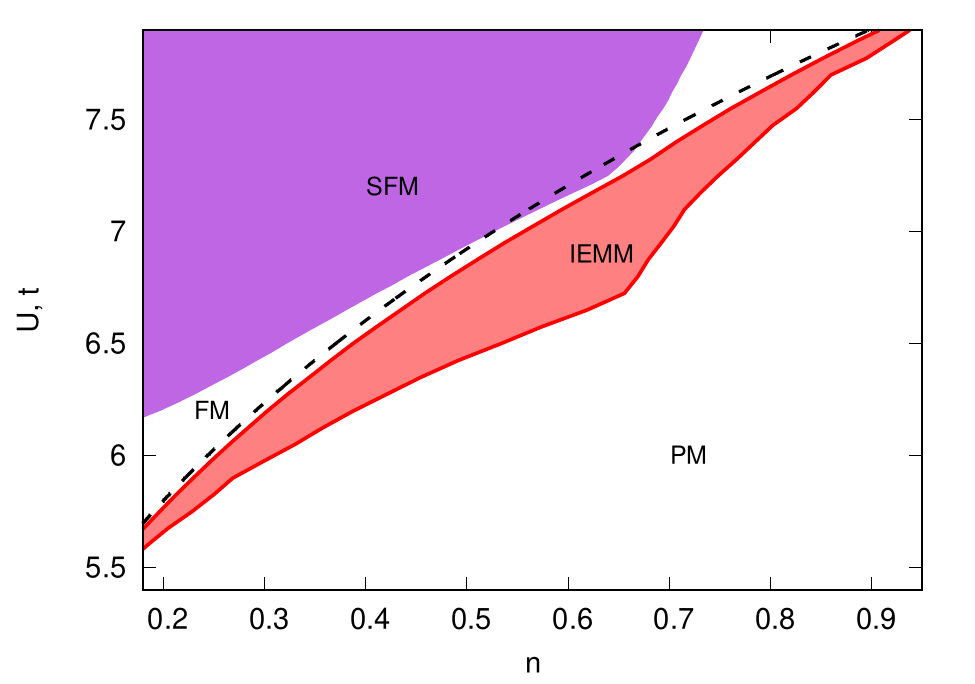}
\caption{Phase diagram in $U-n$ variables for FCC lattice at $\tau = 0.05$ and different temperatures: FM is ferromagnetic region, PM  paramagnetic region. Within the latter, filled (IEMM) areas correspond to metamagnetism regions. The dotted line indicates the curve corresponding the Stoner criterion $U \rho(E_{\rm F})=1$. The SFM region corresponds to the saturated ferromagnetism region.
}
\label{fig:f_d_fcc_tau_0.05}
\end{figure}

The phase diagram $U$---$n$ (see~Fig.~\ref{fig:f_d_fcc_tau_0.05}) has a  behavior, similar to that described earlier for $\tau=0$ (see~Fig.~\ref{fig:f_d_fcc1}), but the filling interval, where IEMM transition occurs, is considerably  reduced for all values of $U$.

However, the values of critical fields are strongly reduced too and become more physically reasonable, see~Fig.~\ref{fig:gap_n_fcc_tau2}.

In this case, saturated magnetism in critical fields is not observed in the whole IEMM transition region and appears only in FM region, being denoted as SFM region in~Fig.~\ref{fig:f_d_fcc_tau_0.05}.

\subsubsection{$\tau \approx -1/2$}
\begin{figure}[h!]
%\begin{minipage}[h]{\linewidth}
\includegraphics[angle=270,width=\linewidth]{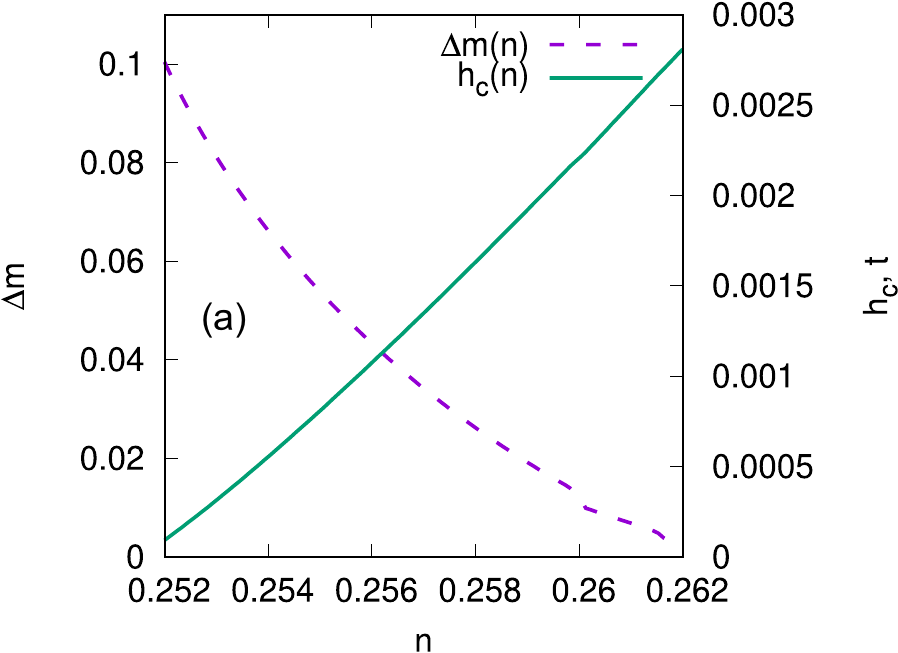}
%\end{minipage}
%\begin{minipage}[h]{\linewidth}
\includegraphics[angle=270,width=\linewidth]{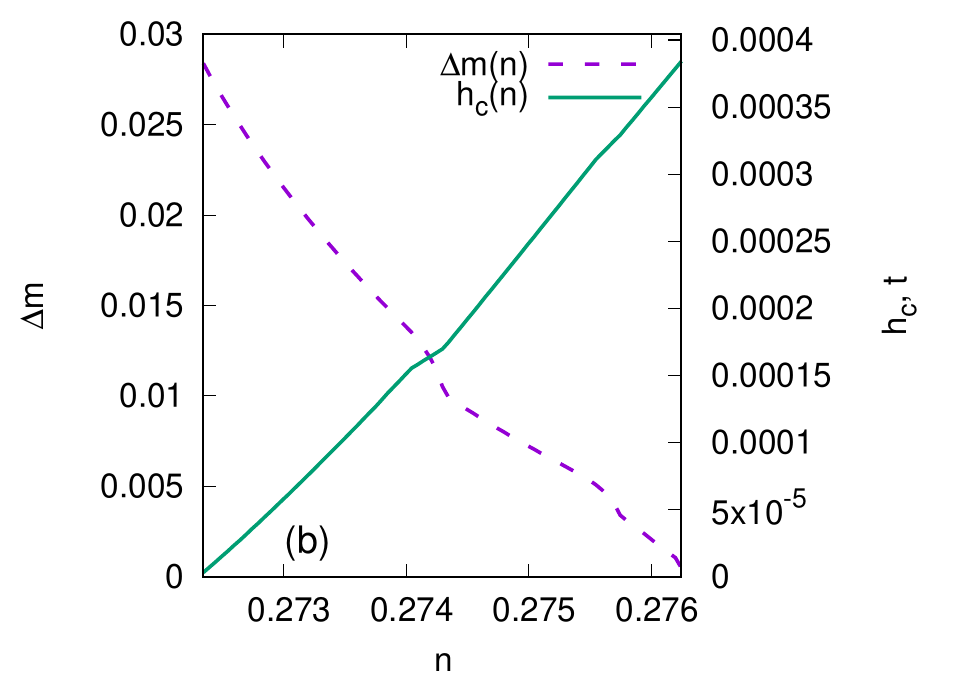}
%\end{minipage}
\caption{The same as for Fig.~\ref{fig:gap_n_rec1} for FCC lattice $U= 1.50 t$, $T= 2.00 \cdot 10^{-4} t$; (a) $\tau = -0.52$; (b) $\tau = -0.54$.
}
\label{fig:hc(n)1}
\end{figure}

In this section, the case of giant vHS  at $\tau' \simeq -1/2$ for FCC lattice is considered, see Sec.~\ref{sec:dos}. 
At weak deviation of $\tau$ from $-1/2$, the giant VSH is transformed into a van Hove plateau.
%IEMM transition is found for values of filling $n$ at which the Fermi level lies to the right of the plateau, where a~considerable curvature takes place (see~Fig.~\ref{fig:dos}). For  other cases (to the left of the plateau and on the plateau), the effect is very weak, i.e.,  occurs in a very small filling range. 
Two values close to $\tau=-1/2$ --- $\tau = -0.52$ and $-0.54$ are chosen, see~Fig.~\ref{fig:dos}. 
In this case, the metamagnetic transition exists in a  small filling interval, see~Fig.~\ref{fig:hc(n)1}. The critical magnetic fields have no anomalously large values, and for $\tau = -0.54$ $h_{\rm c}$ is an order of magnitude smaller than for DOS with $\tau = -0.52$ for the same parameters $U$ and $T$. In such a situation, strong temperature dependences of  critical  field and magnetization jump are expected  (see the corresponding experimental examples in the Introduction).

%Here we study the dependence \bad{$m = m_{\rm i}(n,h_{\rm c}(n))$} for FCC lattice first in the case of low temperatures, $T = 2\cdot10^{-4}t$. %, the bandwidth $W = 18t$ to be 5~eV. 
%\bad{Let us consider the case when the Coulomb interaction parameter $U$ is chosen so that the Stoner criterion is close to being fulfilled (in this case, $U = 1.5t$), then we obtain the magnetic field dependence of the magnetization shown in Fig.~\ref{fig:h_n}(a) for $\tau = -0.52$ and Fig.~\ref{fig:h_n}(b) for $\tau = -0.54$.} 
%\sout{This dependence is typical for a IEMM transition when the magnetization magnetic field dependence exhibits a~jump.} 
%\bad{At filling corresponding to the ferromagnetic phase in the ground state}, there is a noticeable paraprocess, i.e., the dependence of the magnetization on the applied magnetic field. As can be seen from the $m(h)$ dependence for $\tau = -0.54$ in Fig.~\ref{fig:h_n}(b), the critical magnetic field is an order of magnitude smaller than for DOS with $\tau = -0.52$ for the same parameters $U$ and $T$. 
%{\blue[Is this true?] }\bad{This is due to the curvature of $\rho(E)$ at the Fermi level --- the larger the curvature, the smaller the critical magnetic field $h_{\rm c}$.}

%\begin{figure}[h!]
%\includegraphics[angle=270,width=\linewidth]{n_scan.pdf}
%\includegraphics[angle=270,width=\linewidth]{n_scan_tau2.pdf}
%\end{minipage}
%\caption{Dependences of magnetization on magnetic field for FCC lattice at different values of filling $U=1.50 t$, $T =2\cdot 10^{-4} t$. (a) $\tau = -0.52$; (b) $\tau = -0.54$.}
%\label{fig:h_n}
%\end{figure}

\begin{figure}[h!]
\includegraphics[angle=270,width=\linewidth,]{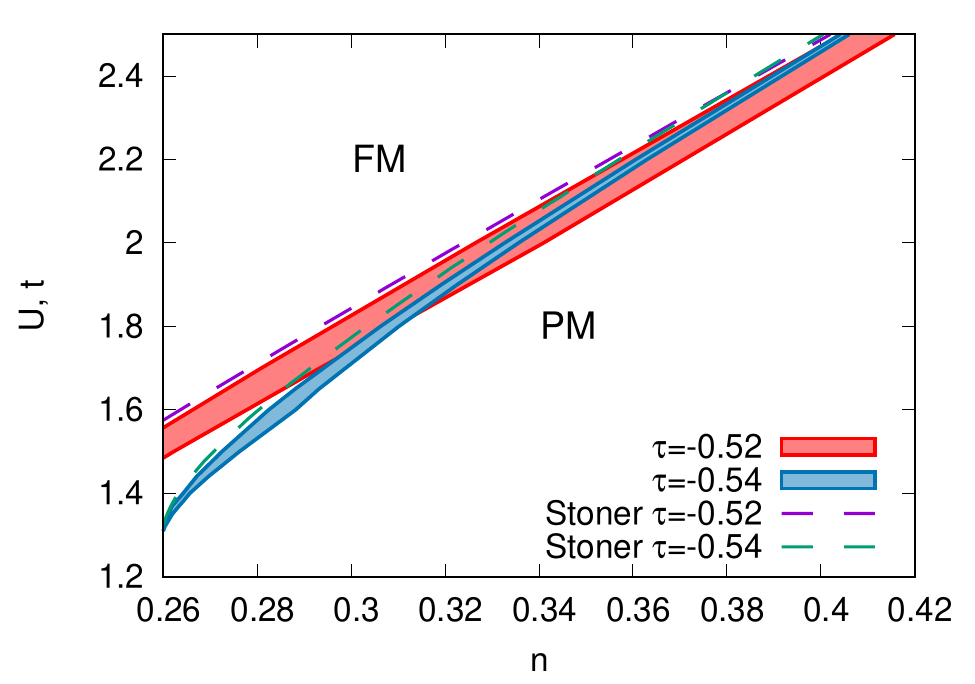}
\caption{Phase diagram for FCC lattice in $U-n$ variables at $T= 2.00 \cdot 10^{-4} t$. The filled regions denote IEMM regions at $\tau = -0.52$ and $-0.54$, the dashed lines indicate the Stoner criterion~(\ref{eq:Stoner}), where $E_{\rm F}$ is taken for the paramagnetic phase for a given $n$.
}
\label{fig:U(n)}
\end{figure}

The region of the IEMM transition follows along the Stoner line, see~Fig.~\ref{fig:U(n)}. The filling interval where the transition is possible remains narrow for any values of $U$. 

\section{Magnetic phase transitions:  Applicability of the Landau theory }\label{sec:Landau}
To explain the occurrence of IEMM transition, the Landau theory is typically used~\cite{shimizu1982itinerant,levitin1988itinerant}. 
In this Section we estimate the validity of the Landau theory for describing IEMM transitions in the Hubbard model. 
The Landau-theory expansion of the free energy has the form
\begin{multline}\label{eq:Landau_free_energy}
F_{\rm Landau}(T,n,h|m) = a_0(T,n) + a_2(T,n) m^2 \\
+ a_4(T,n) m^4 + a_6(T,n) m^6 - h m,  
\end{multline}
where explicit expression for coefficients $a_i$ within HFA are written in Appendix. 
The stability of expansion~\eqref{eq:Landau_free_energy} requires $a_6 > 0$. 
At given $T$, $n$, $h$, a state with $m$ corresponding to global minimum of $F_{\rm Landau}$ with respect to $m$ is established. 
In the case of second-order phase transition which occurs at $a_4>0$, negative (positive) $a_2$ values correspond to FM (PM) phase, and the transition between them ($a_2 = 0$) is continuous. 
Within HFA, the FM state condition $a_2  = (1/4)(\rho^{-1}(E_{\rm F})- U) < 0$ at $T = 0$ produces standard Stoner criterion, see Eq.~\eqref{eq:Stoner}. In the Stoner picture, small negative $a_2$ values correspond to weak itinerant ferromagnetism, whereas small positive $a_2$ to nearly ferromagnetic situation.

Negative $a_4$ can induce first-order magnetic phase transition at $a_2 > 0$. 
In this case, for the zero magnetic field ($h=0$), when the condition $a_2 < a^{\rm II}_{\rm c}$ ($a^{\rm II}_{\rm c} = 3 a_4^2/(5 a_6)$) is satisfied the free energy has two local minima  $m_1=0$ and $m_2 > 0$, and in the case $F_{\rm Landau}(T,n,h=0|m_1) < F_{\rm Landau}(T,n,h=0|m_2)$, a global minimum corresponds to the paramagnetic state. When the magnetic field $h$ is turned on, both minimum levels are lowered proportionally to $m_i$ and at a certain value $h$ the minima will coincide $F_{\rm Landau}(T,n,h|m_1)=F_{\rm Landau}(T,n,h|m_2)$. At this value of magnetic field, the magnetization changes abruptly indicating the IEMM transition. 
For the chosen signs of the coefficients,  ferromagnetic ordering requires fulfillment of the condition $a_2 < a_{\rm c}^{\rm I}$ ($a_{\rm c}^{\rm I}= a_4^2/(4 a_6)$), which follows from $F_{\rm Landau}(T,n,h=0|m_1=0)>F_{\rm Landau}(T,n,h=0|m_2)$. 
In this case, negative value of  
\begin{equation}\label{eq:a4_GS}
a_4(0,n) = \dfrac{1}{4}\dfrac{1}{16 (\rho(E_{\rm F}))^{3}}\left( \dfrac{(\rho'(E_{\rm F}))^2}{\rho^{2}(E_{\rm F})} - \dfrac{\rho''(E_{\rm F})}{3\rho(E_{\rm F})}\right),  
\end{equation}
see explicit expression in Appendix, Eq.~\eqref{eq:coefficients_ai:a4}, 
considerably modifies the criterion of ferromagnetic ordering, thereby invalidating the Stoner criterion, see the phase diagrams.
$a_4$, being squared, should be sufficiently large, which can be achieved owing to vHS (i.e., a large DOS curvature) at the PM phase Fermi level. 
On the basis of the corrected criterion of ferromagnetism ($a_2<a_{\rm c}^{\rm I}$) and the condition for the existence of two local minima of the free energy ($a_2<a_{\rm c}^{\rm II} $) we can write the condition for the IEMM transition
\begin{equation}\label{eq:a2_criterion}
  a_{\rm c}^{\rm I} < a_2 <a_{\rm c}^{\rm II}.
\end{equation}
%where $a_{\rm c}^{\rm I} = a_4^2/(4 a_6)$ and $a_{\rm c}^{\rm II} = 3 a_4^2/(5 a_6)$. 
Since the Landau theory is constructed as an  expansion in magnetization, it is not capable to describe the transition from saturated to  non-saturated FM state. 

%{\blue [Povtor:]}\bad{The region of existence of IEMM transition can be given by two conditions.  The first condition is the presence of $F_{\rm Laudau}''(T,n,h|m) < 0$ interval (a prime denotes $m$ derivative) implying positive curvature $F_{\rm Landau}$ with respect to $m$, so we obtain the condition $\frac{a_2a_6}{a_4^2} < 3/5$. The second condition implies that at $h = 0$, $F_{\rm Landau}(T,n,h=0|m_1)<F_{\rm Landau}(T,n,h=0|m_2)$, where $m_1=0$ and $m_2$ are free energy minima: which directly yields the condition $\frac{a_2a_6}{a_4^2}>\frac{1}{4}$. 
%For occurrence of metamagnetism on the basis of Landau's theory, it is necessary to fulfill the inequality
%Within the Landau theory the condition of IEMM transition reads
%\begin{equation}\label{eq:a2_criterion}
%  a_{\rm c}^{\rm I} < a_2 <a_{\rm c}^{\rm II},
%\end{equation}
%where $a_{\rm c}^{\rm I} = a_4^2/(4 a_6)$ and $a_{\rm c}^{\rm II} = 3 a_4^2/(5 a_6)$. 
%}
 
Note that, generally speaking, higher-order expansion terms cannot be neglected, since they are physically related to the derivatives of bare DOS (see Appendix). Moreover, we will demonstrate that they can become large in the case of closeness of PM phase Fermi level to vHS.

 Describing metamagnetism within the Landau expansion is possible only in the case of small magnetization, many works being devoted to this approach \cite{levitin1988itinerant,shimizu1982itinerant, Yamada1993, goto2001itinerant, yamada2003itinerant, goto1997magnetic, belitz2005tricritical, yamada2007p, berridge2010magnetic}. 
Previously, the expansion of the free energy in powers of magnetization (Landau functional) was obtained in Refs.~{\cite{levitin1988itinerant, shimizu1982itinerant}. In this section, we will compare the results of the Landau theory for  the calculated coefficients $a_2(T,n)$, $a_4(T,n)$, $a_6(T,n)$ with the numerical solution of the Eqs.~(\ref{eq:main_eqiations:n}) and (\ref{eq:main_eqiations:m}), which will allow us to evaluate the limits of applicability of the Landau  functional approach. 

%\subsection{Rectangular lattice}
For example, we consider the case of a rectangular lattice with $\tau_{\rm a}=0.90$, $\tau=0.20$, choosing the filling at which the IEMM transition occurs,  $n=0.62$, see Fig.~\ref{fig:m_h_low_an}. The temperature dependences of coefficients $a_i$ are shown at~Fig.~\ref{fig:F_m_rec}(a). The coefficient $a_6(T,n) > 0$ at  $T > 0.016 t$ and $a_4(T,n) < 0$ at $T < 0.040 t$. That is, the fulfillment of the condition for the $F_{\rm Landau}(T,n|m)$ minimum  is possible at $T>0.016 t$, but at $T>0.040 t$ the coefficient $a_4(T,n)$ becomes positive, and hence the second free energy minimum disappears, so that the IEMM transition is impossible. Comparison of $F_{\rm HFA}$ and $F_{\rm Landau}$ at zero magnetic field $h=0$ and $U=0$ in the temperature region $0.008<T<0.040 t$ Fig.~\ref{fig:F_m_rec}(b) shows good agreement for magnetization values for small $m$ only. %down to $m=0.05$. 
Thus, in this case, where the metamagnetic transition occurs in the region of magnetization $m<0.1$, the results described by the Landau theory should agree well with those obtained from HFA.
\begin{figure}[h!]
\includegraphics[angle=270,width=\linewidth]{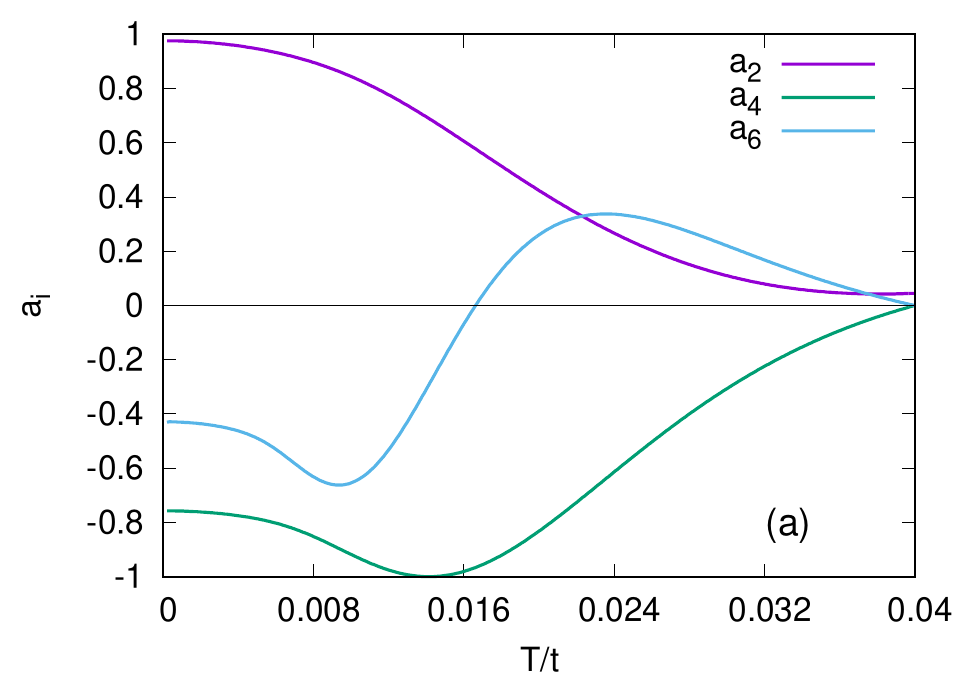}
\includegraphics[angle=270,width=\linewidth]{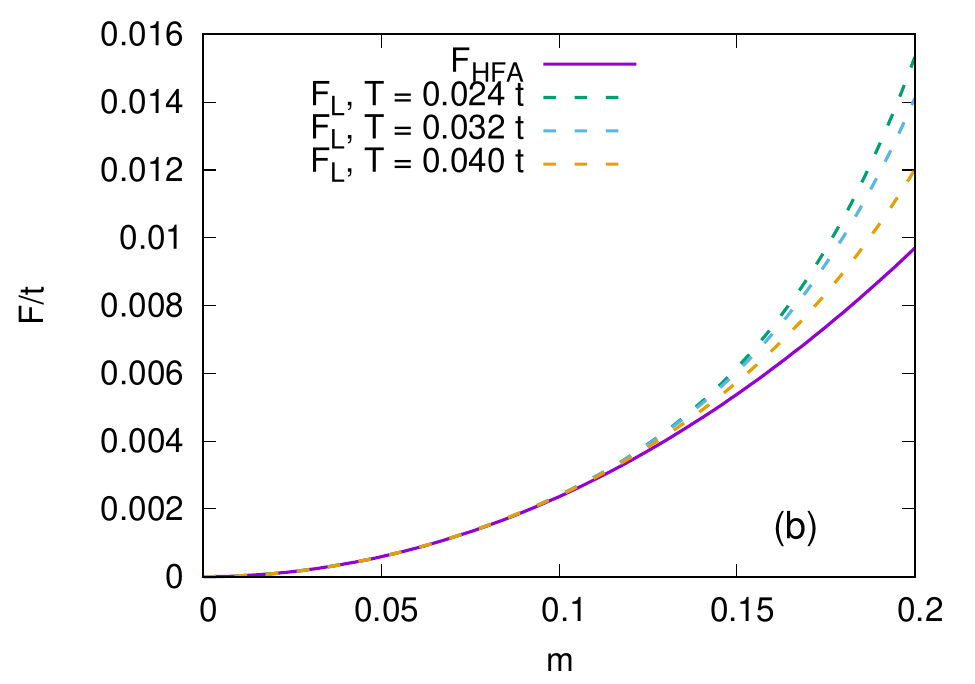}
\caption{ %{\green $[\times2]$}
(a) Temperature dependence of rescaled expansion coefficients $a_{2,4,6}(T,n)$. 
(b)  Dependence of the mean-field free energy $F_{\rm HFA}(T,n|m)$ (solid lines) and Landau free energy $F_{\rm Landau}(T,n|m)$ (dashed lines) on the magnetization $m$ for different temperatures at $U = 0$ for rectangular lattice $\tau_{\rm a}=0.90$, $\tau=0.20$ at $n = 0.62$. 
}
\label{fig:F_m_rec}
\end{figure}
The  $m$ dependence of the free energy within HFA  changes only weakly with temperature, see~Fig.~\ref{fig:F_m_rec}(b), whereas the free energy of the Landau theory has poor convergence to the value of $F_{\rm HFA}(T,n|m)$ at $m>0.12$, which becomes even worse with decreasing temperature. This feature strongly limits using the Landau expansion due to that a necessary number of expansion terms to be retained is {\it a priori} unknown.

The filling and temperature dependence of $a_i$ is shown in Fig.~\ref{fig:a_i_rec}. 
%The sign and value of the expansion coefficients are also strongly influenced by  filling, see Fig.~\ref{fig:a_i_rec}. 
At filling values, which correspond to  PM phase Fermi level coinciding  with vHS, peaks also appear in the $a_i(n)$ dependence. An increase in temperature results in a~decrease of peaks height. The coefficient $a_4(T,n)$ has a negative value at $1.10<n<1.26$ and $T=0.016 t$, which coincides with the region where IEMM transition occurs in in mean-field calculations (see~Fig.~\ref{fig:f_d}).
For the applicability of Landau's theory, it is also necessary for the coefficient $a_6(T,n)$ to be positive (and substantially large) in the region where $a_4(T,n)$ is negative. As we can see from Fig.~\ref{fig:a_i_rec}, $a_6(T,n)$ becomes positive only at temperatures greater than $T=0.016 t$ and in a very small filling region.  
Large $a_i$ values  indicate inapplicability of the expansion which is usually performed in the Landau theory.

%\bad{
%Below we discuss the $n$ dependence of $a_{4,6}$ at $0.55<n<0.63$.  
%In a region where $a_4(T,n)$ has a negative value, the coefficient $a_6(T,n)$ is positive only at the edges or in the middle of this region, but at higher temperatures (Fig.~\ref{fig:a_i_rec}). From the results obtained Fig.~\ref{fig:m_h_low_an}, it can be seen that the IEMM transition occurring at $m<0.05$ corresponds to filling near the edges of the negativity interval ($n=0.55$ and $0.63$) for the coefficient $a_4(T,n)$. Thus, the results obtained for these filling values near  should be in the best agreement, but from Fig.~\ref{fig:m_h_T}(a) one can see that the temperature{\blue???} suppresses IEMM transition most strongly just at these values of $n$.
%}

\begin{figure}[h!]
\includegraphics[angle=270,width=\linewidth]{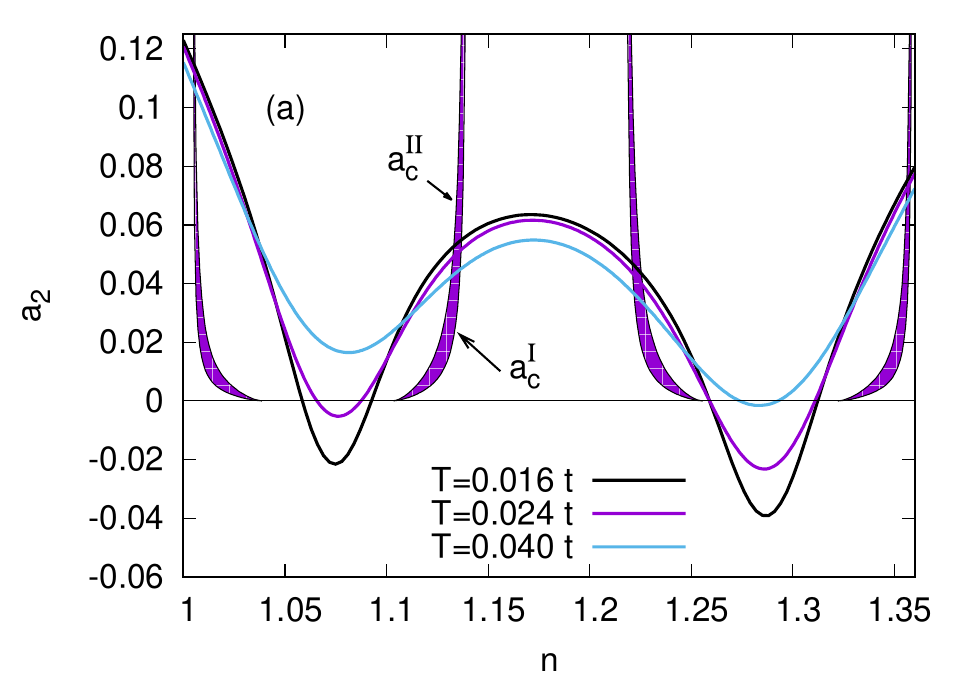}
\includegraphics[angle=270,width=\linewidth]{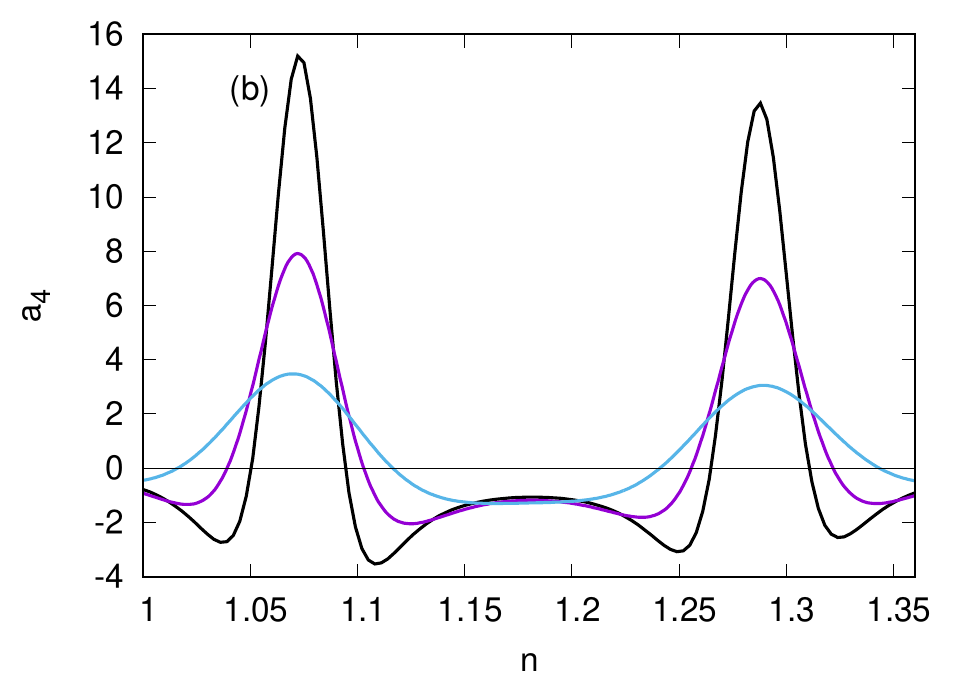}
\includegraphics[angle=270,width=\linewidth]{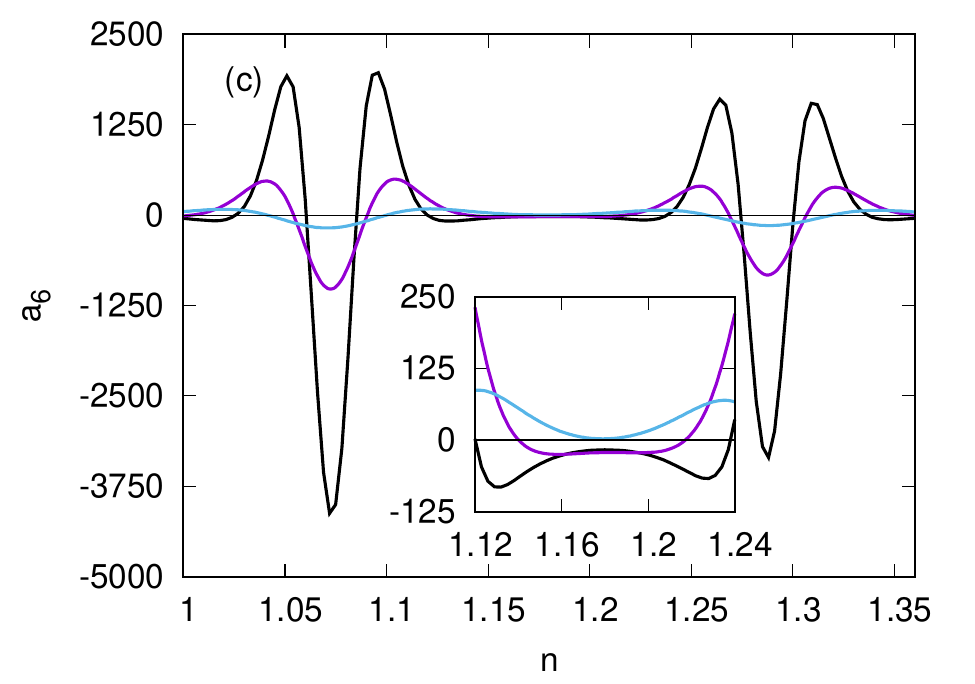}
\caption{%{\green $[\times2]$} 
Dependence on filling of the rescaled expansion coefficients $a_{i}$ for different temperatures for rectangular lattice $\tau_{\rm a}=0.90$, $\tau=0.20$ at $U=3.50 t$. (a) $a_2(n)$, filling indicates region $a_{\rm c}^{\rm I} < a_2 < a_{\rm c}^{\rm II}$ at $T=0.024 t$; (b)   $a_4(n)$; and (c)  $a_6(n)$. 
}
\label{fig:a_i_rec}
\end{figure}

By selecting the parameters corresponding to the conditions $a_2>0$, $a_4<0$, $a_6>0$: $U = 3.70t$, $T=0.032t$, magnetic field dependences were constructed, Fig.~\ref{fig:m_h_land}. The results agree well up to $m=0.04$. At these filling values, metamagnetism is already strongly suppressed by temperature. At $n=1.14$, an IEMM transition occurs at magnetization value for which the results are already poorly consistent.

\begin{figure}[h!]
\includegraphics[angle=270,width=\linewidth]{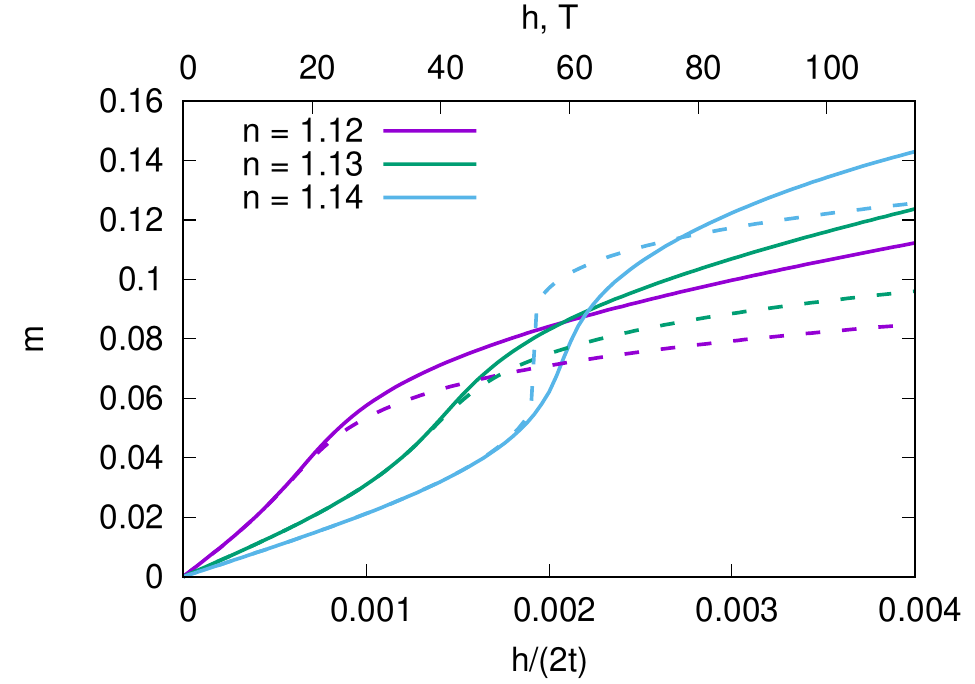}
\caption{%{\green $[\times2]$}
Dependence of magnetization on magnetic field: comparison of numerical calculation with the Landau theory at different filling values  $U=3.70 t$ for rectangular lattice $\tau_{\rm a}=0.90$, $\tau=0.20$ at $T=0.032 t$. The solid line corresponds to calculations from the mean-field approximation, and the dashed line to those from the Landau theory.}
\label{fig:m_h_land}
\end{figure}

\begin{figure}
\includegraphics[angle=270, width=\linewidth]{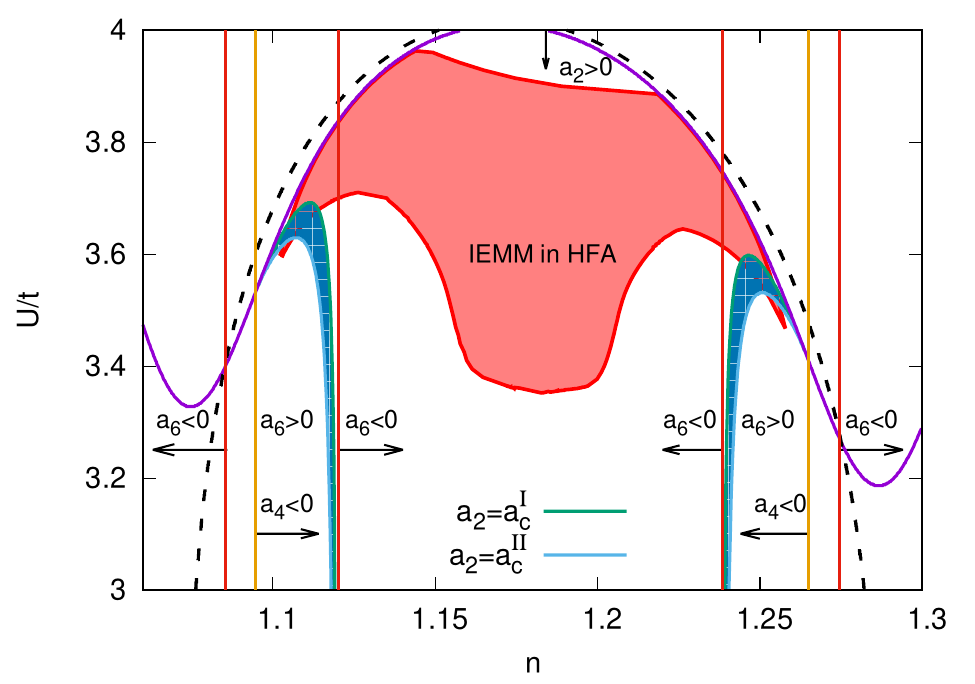}
\caption{\label{fig:Landau_theory_PD}
%{\green $[\times2]$} 
Comparison of phase diagrams in $U-n$ variables obtained from the Landau theory and HFA for a rectangular lattice with $\tau_{\rm a} = 0.9$, $\tau = 0.20$ at $T=0.016 t$. The blue area between the curves $a_2=a_{\rm c}^{\rm I}$ and $a_2=a_{\rm c}^{\rm II}$   corresponds to the region of IEMM transition according to the Landau theory.}
\end{figure}

The Landau theory phase diagram is shown in Fig.~\ref{fig:Landau_theory_PD}. 
One can see that the possibility of the Landau theory to yield correct criterion of IEMM transition is restricted to the region of close vicinity of $a_4 = 0$ line. 
The extension of the validity region requires taking into account more and more terms of the Landau expansion.  
Thus the application of the Landau expansion to describe IEMM transition~\eqref{eq:a2_criterion}~(see Refs.~\onlinecite{levitin1988itinerant,shimizu1982itinerant}) should be performed with a great caution. 

\section{Conclusions}
 As discussed in Introduction, itinerant metamagnetism occurs in the case of large positive DOS curvature.
There are several factors which are favorable for this DOS form: splitting of vHS, reduced space dimensionality  (layered systems), hybridization in the case of degenerate electron bands. We have investigated the possibility of existence of metamagnetism for densities of states possessing peculiarities characteristic for these factors.

The splitting of vHS leads to the appearance of a region with strong positive curvature between vHS peaks, which may favorably affect the occurrence of metamagnetism in a large filling region. We have shown how such a splitting leads to a double-peak structure of the electronic DOS and the formation of a metamagnetic transition between ferromagnetic phases for a two-dimensional rectangular lattice. The metamagnetic transition is preserved by introducing hopping along the $z$~axis. At the same time, the  $U$-region where metamagnetism exists is narrowed owing to smoothing of two-dimensional lattice DOS peaks. However, a considerable metamagnetism region is found for an orthorombic lattice.

 An example of metamagnetism in systems with lowered dimensionality is provided by layered ruthenates Sr$_3$Ru$_2$O$_7$ and Sr$_4$Ru$_3$O$_{10}$ (see  Introduction). 
As for three-dimensional systems, a scenario of metamagnetic transition induced by  position of the Fermi level between two vHS peaks was suggested for metamagnetic transition in MnAs under pressure larger than 14.6 kbar~\cite{1999:Yamada:MnAs}. 

For a FCC lattice with logarithmic divergence at the bottom of the band ($\tau=0$), ferromagnetism has saturated nature, and the IEMM transition occurs in a wide filling region, but in anomalously strong magnetic fields. When considering the case  $\tau=0.05$, the logarithmic DOS divergence becomes smeared, and the critical magnetic fields of the IEMM transition become more realistic.

For the  FCC lattice with $\tau \simeq - 0.5$, metamagnetism arises at the position of the Fermi level near single vHS in the form of a plateau, in the region with a positive curvature. However, as reflected in the presented phase diagrams, the filling interval of existence of metamagnetism is not wide. Therefore we can conclude that  the DOS structure  is insufficient to observe the effect in a larger region for such a case. However, in this case we can expect low values of critical fields and strong temperature dependence of the magnetization jump. Further investigation of this dependence in the presence of vHS, as well as of non-trivial temperature dependence of  magnetic susceptibility in the paramagnetic phase, would be of interest.

%(Since the mean-field approximation used seems to be unjustified?? for the results?? given, one can state that phase diagram is only qualitatively correct. -
%chereschur sanokritichno!! luchshe napisat' o perenermirovke kriteriya Stonera):

We have performed our consideration within the mean-field approximation. Unlike the finite-temperature case, this approach is, generally speaking, satisfactory at $T=0$ (e.g., in the Moriya theory \cite{Moriya_book} the ground state is described by the Stoner theory, but spin fluctuations are important at finite $T$). However, in the presence of vHS the ground-state and low temperature renormalizations are expected to have strong parameter dependence. In particular, the treatment within functional renormalization group for 3D spectrum with giant vHS shows that in the absence of enhanced incommensurate magnetic fluctuation the Hubbard interaction $U$ can be considered as a Coulomb interaction parameter being renormalized by scattering in particle-particle channel~\cite{igoshev2023ferromagnetic}. Thus we expect that correlation effects result in renormalization of the parameter~$U$ only. 
Other related cases are provided by the 2D situation with $t'\lesssim t'/2$  and  systems with noticeable Fermi surface feature like nesting~\cite{2003:Katanin,2007:Igoshev,2011:Igoshev,2017:Stepanenko}, where long-wave incommensurate fluctuations and corresponding scattering channel dominate.

On one hand, we propose pure qualitative reason for IEMM transition - pair of vHS singularity or vHS plateau providing high-curvature energy region of electron density of states. 
One the other hand, taking into account of spin fluctuation within some Moriya's-like approach should increase critical $U$ for IEMM transition and only some quantitative change of the phase diagram. 
Rather large critical $U$ obtained for metamagnetic transition suggests that full-fledged theory of IEMM caused by vHS should concern both many-orbital effects and Hund's interaction compensating insufficiency of direct Coulomb interaction. 
A possible correlation-induced increase of critical $U$ can be compensated by Hund's interaction in some realistic multiband case, which can favor ferromagnetic ordering as well as IEMM transition.

Our study of the applicability of Landau's theory for metamagnetism has shown that an adequate description of this phenomenon is possible only when considering the terms in the expansion up to the eighth order and at high enough temperature , which strongly reduces the limits of applicability of Landau's theory to describe this phenomenon.

\section{Acknowledgments}
We are grateful to N.~V.~Mushnikov for fruitful discussions. 
The studies are partially supported by the state assignment of the Ministry of Science and Higher Education of the Russian Federation (theme ``Quant'' No. 122021000038-7) and by the Development Program of the Ural
Federal University within the strategic academic leadership program ``Priority-2030''. 
Numerical computations were performed on the Uran supercomputer at the IMM UB RAS. 

\appendix
\section{Landau expansion derivation}\label{app:expansion}

 The coefficients $a_2$, $a_4$, $a_6$ in the free energy expansion \eqref{eq:F_HFA} can be calculated from 
 \begin{eqnarray}\label{eq:coefficients_ai:a2}
       a_2 &=& \dfrac{1}{4} (\Pi_{0}^{-1} - U ),\\
\label{eq:coefficients_ai:a4}
       a_4 &=& \dfrac{1}{4}\dfrac{1}{16 \Pi_{0}^{3}}\left( \dfrac{\Pi_{0}^{'2}}{\Pi_{0}^{2}} - \dfrac{\Pi_{0}''}{3\Pi_{0}}\right),
\end{eqnarray}
\begin{multline}\label{eq:coefficients_ai:a6}
a_6 = \dfrac{1}{4} \dfrac{1}{16} \dfrac{1}{24 \Pi_{0}^{5}}\left(7 \left(\dfrac{\Pi_{0}'}{\Pi_{0}} \right)^4 - 7\dfrac{\Pi_{0}^{'2}}{\Pi_{0}^{3}} \Pi_{0}'' \right. \\
+\left. \dfrac{2}{3}\left(\dfrac{\Pi_{0}''}{\Pi_{0}} \right)^2 + \dfrac{\Pi_{0}'\Pi_{0}^{'''}}{\Pi_{0}^{2}} - \dfrac{1}{15}\dfrac{\Pi_{0}''''}{\Pi_{0}}\right),
\end{multline}
\begin{multline}
    a_8 = \frac{1}{128\Pi_{0}^{13}}\left[17325\Pi_{0}\Pi_{0}'^4\Pi_{0}'' \right. \\
+126\Pi_{0}^2\Pi_{0}'^2 \left(3\Pi_{0}\Pi_{0}^{(4)}-50\Pi_{0}''^2\right)\\
-28\Pi_{0}^3\Pi_{0}'\left(\Pi_{0}\Pi_{0}^{(5)}-45\Pi_{0}^{(3)}\Pi_{0}''\right)\\
- 10395\Pi_{0}'^6-3150\Pi_{0}^2\Pi_{0}^{(3)}\Pi_{0}'^3\\
+\Pi_{0}^3 \left(280\Pi_{0}''^3-56\Pi_{0}\Pi_{0}^{(4)}\Pi_{0}'' \right.\\
\left. \left. +\Pi_{0} \left(\Pi_{0}\Pi_{0}^{(6)}-35(\Pi_{0}^{(3)})^2\right)\right) \right],
\end{multline}
where zero wave vector polarization operator reads
\begin{equation}
    \Pi_0 = \dfrac{1}{TN}\sum \limits_{{\bf k}} \zeta\left(\dfrac{\epsilon_{\bf k} - E_{\rm F}}{T}\right)
\end{equation}
with $\zeta(x) = \dfrac{1}{4\cosh^{2}\frac{x}{2}}$, primes denote derivatives with respect to $E_{\rm F}$ argument.

%\bibliography{references}

\end{document}